\documentclass[onecolumn]{IEEEtran}
\usepackage[mathscr]{eucal}
\usepackage[cmex10]{amsmath}
\usepackage{epsfig,epsf,psfrag}
\usepackage{amssymb,amsmath,amsthm,amsfonts,latexsym}
\usepackage{amsmath,graphicx,bm,xcolor,url,overpic}
\usepackage{fixltx2e}
\usepackage{array}
\usepackage{verbatim}
\usepackage{bm}
\usepackage{algorithmic}
\usepackage{algorithm}
\usepackage{verbatim}
\usepackage{textcomp}
\usepackage{mathrsfs}
\usepackage{epstopdf}

\newcommand{\openone}{\leavevmode\hbox{\small1\normalsize\kern-.33em1}}

\catcode`~=11 \def\UrlSpecials{\do\~{\kern -.15em\lower .7ex\hbox{~}\kern .04em}} \catcode`~=13 

\allowdisplaybreaks[4]

\newcommand{\nn}{\nonumber}

\newcommand{\calN}{\mathcal{N}}

\newcommand{\calP}{\mathcal{P}}
\newcommand{\calQ}{\mathcal{Q}}

\newcommand{\calX}{\mathcal{X}}

\newcommand{\rmd}{\mathrm{d}}

\newcommand{\rme}{\mathrm{e}}

\newcommand{\rmP}{\mathrm{P}}

\newcommand{\rmQ}{\mathrm{Q}}

\newcommand{\rmV}{\mathrm{V}}

\newcommand{\bbE}{\mathsf{E}}

\newcommand{\bbN}{\mathbb{N}}

\newcommand{\bbR}{\mathbb{R}}

\DeclareMathAlphabet{\mathbsf}{OT1}{cmss}{bx}{n}
\DeclareMathAlphabet{\mathssf}{OT1}{cmss}{m}{sl}

\DeclareSymbolFont{bsfletters}{OT1}{cmss}{bx}{n}  
\DeclareSymbolFont{ssfletters}{OT1}{cmss}{m}{n}
\DeclareMathSymbol{\bsfGamma}{0}{bsfletters}{'000}
\DeclareMathSymbol{\ssfGamma}{0}{ssfletters}{'000}
\DeclareMathSymbol{\bsfDelta}{0}{bsfletters}{'001}
\DeclareMathSymbol{\ssfDelta}{0}{ssfletters}{'001}
\DeclareMathSymbol{\bsfTheta}{0}{bsfletters}{'002}
\DeclareMathSymbol{\ssfTheta}{0}{ssfletters}{'002}
\DeclareMathSymbol{\bsfLambda}{0}{bsfletters}{'003}
\DeclareMathSymbol{\ssfLambda}{0}{ssfletters}{'003}
\DeclareMathSymbol{\bsfXi}{0}{bsfletters}{'004}
\DeclareMathSymbol{\ssfXi}{0}{ssfletters}{'004}
\DeclareMathSymbol{\bsfPi}{0}{bsfletters}{'005}
\DeclareMathSymbol{\ssfPi}{0}{ssfletters}{'005}
\DeclareMathSymbol{\bsfSigma}{0}{bsfletters}{'006}
\DeclareMathSymbol{\ssfSigma}{0}{ssfletters}{'006}
\DeclareMathSymbol{\bsfUpsilon}{0}{bsfletters}{'007}
\DeclareMathSymbol{\ssfUpsilon}{0}{ssfletters}{'007}
\DeclareMathSymbol{\bsfPhi}{0}{bsfletters}{'010}
\DeclareMathSymbol{\ssfPhi}{0}{ssfletters}{'010}
\DeclareMathSymbol{\bsfPsi}{0}{bsfletters}{'011}
\DeclareMathSymbol{\ssfPsi}{0}{ssfletters}{'011}
\DeclareMathSymbol{\bsfOmega}{0}{bsfletters}{'012}
\DeclareMathSymbol{\ssfOmega}{0}{ssfletters}{'012}

\DeclareMathOperator*{\argmin}{arg\,min}

\newtheorem{theorem}{Theorem} 
\newtheorem{lemma}[theorem]{Lemma}

\newtheorem{definition}{Definition}

\usepackage{cite}
\usepackage{subfigure,epstopdf} 

\graphicspath{{./figures/}}
\begin{document}

\title{Refined Asymptotics for Rate-Distortion using Gaussian Codebooks for Arbitrary Sources}
\author{\IEEEauthorblockN{Lin Zhou, Vincent Y.~F.~Tan and Mehul Motani}  \thanks{The authors are with the Department of Electrical and Computer Engineering, National University of Singapore (Emails: lzhou@u.nus.edu, vtan@nus.edu.sg, motani@nus.edu.sg). }
}
\maketitle

\begin{abstract}
The rate-distortion saddle-point problem considered by Lapidoth (1997) consists in finding the minimum rate to compress an arbitrary ergodic source when one is constrained to use a random Gaussian codebook and minimum (Euclidean) distance encoding is employed.  We extend Lapidoth's analysis in several directions in this paper. Firstly, we consider refined asymptotics. In particular, when the source is stationary and memoryless, we establish the second-order, moderate, and large deviation asymptotics of the problem. Secondly, by ``random Gaussian codebook'', Lapidoth referred to a collection of random codewords, each of which is drawn independently and uniformly from the surface of an $n$-dimensional sphere. To be more precise, we term this as a spherical codebook. We also consider i.i.d.\ Gaussian codebooks in which each random codeword is drawn independently from a product Gaussian distribution. We derive the second-order, moderate, and large deviation asymptotics when i.i.d.\ Gaussian codebooks  are employed.  In contrast to the recent work on the channel coding counterpart by Scarlett, Tan and Durisi (2017), the dispersions for spherical and i.i.d.\ Gaussian codebooks are identical. The ensemble excess-distortion exponents for both spherical and i.i.d.\ Gaussian codebooks are established for all rates. Furthermore, we show that the i.i.d.\ Gaussian codebook has a strictly larger excess-distortion exponent than its spherical counterpart for any rate greater than the ensemble rate-distortion function derived by Lapidoth.
\end{abstract}

\begin{IEEEkeywords}
Lossy data compression, Rate-distortion, Gaussian codebook, Mismatched encoding, Minimum distance encoding,  Ensemble tightness, Second-order asymptotics, Dispersion, Moderate deviations, Large deviations
\end{IEEEkeywords}

\section{Introduction}

In the traditional lossy data compression problem~\cite[Section 3.6]{el2011network}, one seeks to find the minimum rate of compression of a source while allowing it to be reconstructed to within a distortion  $D$ at the output of the decompressor. Shannon~\cite{shannon1959coding} established the rate-distortion function for stationary and memoryless sources. However, practical considerations on the system design often necessitate a particular encoding strategy. This then constitutes a {\em mismatch} problem in which the codebook is optimized for a source with one distribution but used to compress a source of a {\em different} distribution. For example, one might be interested to use a {\em Gaussian codebook}---a codebook that is optimal for a memoryless Gaussian  source---to compress a source that is arbitrary. For all ergodic sources with second moment $\sigma^2$, Lapidoth~\cite[Theorem~3]{lapidoth1997} established that the (ensemble) rate-distortion function is
\begin{equation}
R_{\sigma^2}^*(D)=\frac{1}{2}\log\max\bigg\{ 1,\frac{\sigma^2}{D}\bigg\}. \label{eqn:rdf} 
\end{equation}
The term ``Gaussian codebook'' requires some qualifications; Lapidoth~\cite{lapidoth1997} used this term to refer to a collection of random codewords each of which is drawn independently and uniformly from the surface of a sphere in $n$-dimensions. In this work, we  term this random codebook as a {\em  spherical codebook} and, for the sake of comparison, we also consider {\em i.i.d.\ Gaussian codebooks} in which each component of each codeword is drawn independently from a (univariate) Gaussian distribution. In the spirit of recent emphases on refined asymptotics that bring to light the tradeoff between the coding rate, the blocklength, and the probability of excess-distortion, in this paper, we establish ensemble-tight second-order coding rates, moderate deviations constants and excess-distortion exponents. 

\subsection{Main Contributions and Related Works}
Our main contributions are as follows:
\begin{enumerate}
\item We conduct a second-order asymptotic analysis~\cite{strassen1962asymptotische, hayashi2008source, polyanskiy2010finite} for the rate-distortion saddle-point problem for stationary and memoryless sources that satisfy certain technical conditions. Here, the probability of excess-distortion is allowed to be non-vanishing and the spotlight is shone on the additional rate, above the rate-distortion function, required at finite blocklengths to compress the source to within the  prescribed probability of excess-distortion. This work complements that of Kostina and Verd\'u~\cite{kostina2012fixed} and Ingber and Kochman~\cite{ingber2011dispersion} who established the second-order asymptotics (or dispersion) for compressing (discrete and Gaussian) memoryless sources when the encoder is unconstrained. 
We show  that the {\em mismatched dispersions} (and the first-order coding rates or rate-distortion functions) for spherical and i.i.d.\ Gaussian codebooks are identical; this implies that there is no performance loss in terms of the backoff from the first-order fundamental limit regardless of which type of Gaussian codebook one uses. This is in contrast to the recent work by   Scarlett, Tan, and Durisi~\cite{scarlett2017mismatch} on the channel coding counterpart of this problem~\cite{lapidoth1996}. It was shown in \cite{scarlett2017mismatch} that the dispersions for both types of codebooks are different. For the lossy source coding case, the dispersions are common and depend  only on the second and fourth moments of the source through a simple formula. We provide intuition for why this is the case after the statement of  Theorem~\ref{secondorder}. We recover the dispersion of lossy compression of  Gaussian memoryless sources (GMSes)~\cite{kostina2012fixed,ingber2011dispersion} by particularizing the arbitrary source to be a  Gaussian.

\item Next,  we conduct  moderate deviations analysis \cite{altugwagner2014,polyanskiy2010channel} for the same problem under an additional assumption on the source. Here, the rate of the codebook approaches the rate-distortion function at a speed slower than the reciprocal of the square root of the blocklength. One then seeks the subexponential rate of decay of the  probability of excess-distortion. This analysis complements that of Tan~\cite{tan2012moderate} who considered the unconstrained encoding case for (discrete and Gaussian) memoryless sources. This was generalized to the successive refinement problem by the present authors~\cite{zhou2016second}. We again show that the moderate deviations constants are identical and that for GMSes can be easily recovered.

\item Finally, we consider the large deviations regime~\cite{csiszar2011information,gallagerIT} in which the rate of the codebook is constrained to be strictly above the rate-distortion function and one seeks to establish the exponential rate of decay of the probability of excess-distortion. Our analysis complements that of Ihara and Kubo~\cite{ihara2000error} who used ideas from Marton~\cite{Marton74} to find the excess-distortion exponent for compressing a GMS.  We establish the ensemble excess-distortion exponents for both spherical and i.i.d.\  Gaussian codebooks. We recover the excess-distortion exponent of lossy compression of GMSes~\cite{ihara2000error} by particularizing the source to be a Gaussian when an i.i.d.\ Gaussian codebook is used. Furthermore, we show that the i.i.d.\ Gaussian codebook has a strictly larger excess-distortion exponent than its spherical counterpart for any  rate $R>R^*_{\sigma^2}(D)$. We  illustrate this result using two numerical examples.
\end{enumerate}

\subsection{Organization of the Rest of the Paper}
The rest of the paper is organized as follows. We set up the notation, formulate our problem precisely, and present existing results in Section \ref{sec:model}. In Section \ref{sec:mainresults}, we present our main results. These include results concerning second-order, moderate, and large deviation  asymptotics.  Sections \ref{sec:proofsecondorder} to \ref{sec:proofld} are devoted to the proofs for each of these asymptotic results respectively.

\section{The Rate-Distortion Saddle-Point Problem}
\label{sec:model}
\subsection{Notation}
\label{sec:notation}
Random variables and their realizations are in upper  (e.g.,\ $X$) and lower case (e.g.,\ $x$) respectively. All sets are denoted in calligraphic font (e.g.,\ $\mathcal{X}$). We use $\bbR$, $\bbR_+$, and $\bbN$ to denote the set of real numbers, non-negative real numbers, and  natural numbers respectively.  For any two natural numbers $a$ and $b$ we use $[a:b]$ to denote the set of all natural numbers   between $a$ and $b$ (inclusive). We use $\exp\{x\}$ to denote $e^x$ and $\lceil x\rceil$ to denote the smallest integer greater than $x$. All logarithms are base $e$. We use $\rmQ(\cdot)$ to denote the standard Gaussian complementary cumulative distribution function (cdf) and  $\rmQ^{-1}(\cdot )$ its inverse. For any random variable $X$, we use $\Lambda_X(\theta)$ to denote the cumulant generating function $\log \bbE[\exp \{\theta X\}]$ (where $\theta\in\bbR$). We use $\Lambda_X^*(t)$ (where $t\in\bbR$) to denote the Fenchel-Legendre transform (convex conjugate) of the cumulant generating function, i.e., $\sup_{\theta \geq 0} \{\theta t-\Lambda_X(\theta)\}$.  Let $X^n:=(X_1,\ldots,X_n)$ be a random vector of length $n$ and $x^n=(x_1,\ldots,x_n)$ is a particular realization. We use $\|x^n\|=\sqrt{\sum_i x_i^2}$ to denote the $\ell_2$ norm of a vector $x^n\in\bbR^n$. Given two sequences $x^n$ and $y^n$, the quadratic distortion measure (squared Euclidean norm) is defined as $d(x^n,y^n):=\frac{1}{n}\|x^n-y^n\|^2=\frac{1}{n}\sum_{i=1}^n(x_i-y_i)^2$. For any two sequences $\{a_n\}_{n\geq 1}$ and $\{b_n\}_{n\geq 1}$, we write $a_n\sim b_n$ to mean $\lim_{n\to\infty}{a_n} / {b_n}=1$.

\subsection{System Model}
Consider arbitrary source $X$ with distribution  (probability mass function or probability density function) $f_X$  satisfying
\begin{align}
\bbE [X^2]=\sigma^2,~\zeta:=\bbE [X^4]<\infty,~\bbE[X^6]<\infty\label{sourceconstraint}.
\end{align} 
In this paper, we consider memoryless sources and thus $X^n$ is an i.i.d.\ sequence where each component is generated according to $f_X$. We consider the rate-distortion saddle-point problem~\cite[Theorem 3]{lapidoth1997} with an admissible distortion level $0<D<\sigma^2$. This is the lossy source coding problem~\cite[Section~3.6]{el2011network} where one is constrained to use random Gaussian codebooks (spherical or i.i.d.) and an encoding strategy which chooses the codeword that minimizes the quadratic distortion measure.

\begin{definition}
An $(n,M)$-code for the rate-distortion saddle-point problem consists of 
\begin{itemize}
\item A set of $M$ codewords $\{Y^n(i)\}_{i=1}^M$ known by both the encoder and decoder;
\item An encoder $f$ which maps the source sequence $X^n$ into the index of the codeword that minimizes the quadratic distortion with respect to the source sequence $X^n$, i.e.,
\begin{align}
f(X^n)
&:=\argmin_{i\in[1:M]} d\big(X^n,Y^n(i) \big).
\end{align}
\item A decoder $\phi$ which declares the reproduced sequence as the codeword with index $f(X^n)$, i.e.,
\begin{align}
\phi(f(X^n))=Y^n(f(X^n)).
\end{align}
\end{itemize}
\end{definition}

Throughout the paper, we consider random Gaussian codebooks. To be specific, we consider two types of Gaussian codebooks. 
\begin{itemize}
\item  First, we consider the {\em spherical codebook} where each codeword $Y^n$ is generated independently and uniformly over a sphere with radius $\sqrt{n(\sigma^2-D)}$, i.e.,
\begin{align}
Y^n\sim f_{Y^n}^{\rm{sp}}(y^n)=\frac{1\{\|y^n\|^2-n(\sigma^2-D)\}}{S_n(\sqrt{n(\sigma^2-D) })}\label{spherecodebook},
\end{align}
where $1\{\cdot\}$ is the indicator function, $S_n(r)= {n\pi^{n/2}} r^{n-1} / \Gamma(\frac{n+2}{2})$ is the surface area of an $n$-dimensional sphere with radius $r$, and $\Gamma(\cdot)$ is the Gamma function. For GMSes, the spherical codebook is second-order optimal~(cf.~\cite[Theorem~40]{kostina2012fixed}).
\item  Second, we consider the {\em  i.i.d.\  Gaussian codebook} where each codeword $Y^n$ is generated independently according to the following product Gaussian distribution with variance $\sigma^2-D$, i.e.,
\begin{align}
Y^n\sim f_{Y^n}^{\rm{iid}}(y^n)=\prod_{i=1}^n\frac{1}{\sqrt{2\pi(\sigma^2-D)}}\exp\bigg\{-\frac{y_i^2}{2(\sigma^2-D)}\bigg\}\label{iidcodebook}.
\end{align}
The i.i.d.\ Gaussian  codebook is also second-order optimal~(cf.\ \cite[Theorem 12]{kostina2012fixed}) for a GMS.
\end{itemize}


The {\em (ensemble) excess-distortion probability} with $M$ codewords is defined as
\begin{align}
\rmP_{\rme,n}(M)
&:=\Pr\{d(X^n,\phi(f(X^n)))>D\}\label{def:excessdp}\\
&=\bbE_{f_X^n} \left[\big(1-\Pr\{d(X^n,Y^n)\leq D|\,X^n\}\big)^M\right]\label{peforboth},
\end{align}
where \eqref{peforboth} follows from \cite[Theorem 9]{kostina2012fixed} and the inner probability (over $Y^n$ which is independent of $X^n$) is calculated either with respect to the right hand side of~\eqref{spherecodebook} if we use a spherical codebook or the right hand side of \eqref{iidcodebook} if we use an i.i.d.\  Gaussian codebook. Note that the probability in \eqref{def:excessdp} is averaged over the source {\em as well as the  random codebook}. This is in contrast to the traditional lossy source coding analysis~\cite{Marton74,kostina2012fixed} where the excess-distortion probability is averaged over the source only. The additional average over the codebook allows us to pose questions concerning  ensemble tightness in the spirit of~\cite{gallager_ensemble,scarlett2017mismatch}.

\subsection{Existing Results and Definitions}

Let $M_{\rm{sp}}^*(n,\varepsilon,\sigma^2,D)$ be the minimum number of codewords required to compress a length-$n$ source sequence so that the excess-distortion probability with respect to   distortion level $D$ is no larger than $\varepsilon\in(0,1)$ when a spherical  codebook is used. Similarly, let $M_{\rm{iid}}^*(n,\varepsilon,\sigma^2,D)$ be the corresponding quantity when an i.i.d.\  Gaussian codebook is used. Lapidoth~\cite[Theorem 3]{lapidoth1997} showed that for any ergodic source with finite second moment $\sigma^2$ and any $\varepsilon\in(0,1)$, 
\begin{align}
\lim_{n\to\infty}\frac{1}{n}\log M_{\rm{sp}}^*(n,\varepsilon,\sigma^2,D)=\frac{1}{2}\log\frac{\sigma^2}{D} \quad\mbox{nats per source symbol}.
\end{align}
As we will show via a by-product of Theorem~\ref{secondorder}, for any source satisfying \eqref{sourceconstraint} and any $\varepsilon\in(0,1)$, we also have
\begin{align}
\lim_{n\to\infty}\frac{1}{n}\log M_{\rm{iid}}^*(n,\varepsilon,\sigma^2,D)=\frac{1}{2}\log\frac{\sigma^2}{D} \quad\mbox{nats per source symbol} \label{iidfotrue}.
\end{align}

In this paper, we are interested in second-order, large, and moderate deviations analyses. These analyses provide a refined understanding of  the tradeoff between the rate, the blocklength and the excess-distortion probability. In the study of second-order asymptotics, a non-vanishing excess-distortion probability is allowed and we aim to find the back-off from the first-order coding rate (the rate-distortion function) $R^*(\sigma^2,D)=\frac{1}{2}\log\frac{\sigma^2}{D}$.
\begin{definition}
\label{def:secondorder}
Fix any $\varepsilon\in [0,1)$. The spherical second-order coding rate is defined as
\begin{align}
L^*_{\rm{sp}}(\varepsilon)
&:=\limsup_{n\to\infty}\frac{1}{\sqrt{n}}\left(\log M_{\rm{sp}}^*(n,\varepsilon,\sigma^2,D)-R^*(\sigma^2,D)\right).
\end{align}
Similarly, we define the i.i.d.\ second-order coding rate as $L^*_{\rm{iid}}(\varepsilon)$. 
\end{definition}

The moderate deviations regime interpolates between the large deviations (cf.\ Definition \ref{def:ee} to follow) and the second-order regimes. In this regime, we are interested in a sequence of $(n,M)$-codes whose rates approach the first-order coding rate $R^*(\sigma^2,D)=\frac{1}{2}\log\frac{\sigma^2}{D}$ and whose excess-distortion probabilities vanish simultaneously.

\begin{definition}
\label{def:md}
Consider any sequence $\{\xi_n\}_{n\in\bbN}$ such that as $n\to\infty$
\begin{align}
\xi_n\to 0 \quad
\mathrm{and}\quad\sqrt{\frac{n}{\log n}}\xi_n\to\infty\label{req:xin}.
\end{align}
The spherical moderate deviations constant is defined as
\begin{align}
\nu^*_{\rm{sp}}
&:=\liminf_{n\to\infty}-\frac{1}{n\xi_n^2}\log \rmP_{\rme,n}\big(\big\lceil\exp\big(n(R^*(\sigma^2,D)+\xi_n )\big) \big\rceil\big)
\end{align}
Similarly, we define the i.i.d.\ moderate deviations constant as $\nu^*_{\rm{iid}}$.
\end{definition}

In the large deviations regime, we characterize the speed of the exponential decay of the excess-distortion probability for codes with a rate upper bounded by $R$.
\begin{definition}
\label{def:ee}
The rate-$R$ spherical excess-distortion exponent is defined as
\begin{align}
E^*_{\rm{sp}}(R)
:=\liminf_{n\to\infty}-\frac{1}{n\xi_n^2}\log \rmP_{\rme,n}\big(\big\lceil\exp (nR )\big\rceil\big).
\end{align}
Similarly, we define the rate-$R$ i.i.d.\ excess-distortion exponent $E^*_{\rm{iid}}(R)$.
\end{definition}

\section{Main Results}
\label{sec:mainresults}
\subsection{Second-Order Asymptotics}
Our first result pertains to the second-order coding rate. Recall the definitions of $\sigma^2$ and $\zeta$ in \eqref{sourceconstraint}. Let the {\em mismatched dispersion} be defined as
\begin{align}
\rmV(\sigma^2,\zeta)&:=\frac{\zeta-\sigma^4}{4\,\sigma^4} =  \frac{{\rm{Var}}[X^2]}{ 4 \, (  \bbE[X^2]^2) }.\label{def:rmvsphere}
\end{align}
\begin{theorem}
\label{secondorder}
Consider an arbitrary memoryless source $X$ satisfying \eqref{sourceconstraint}. For any $\varepsilon\in [0,1)$, 
\begin{align}
L^*_{\rm{sp}}(\varepsilon)=L^*_{\rm{iid}}(\varepsilon)=\sqrt{\rmV(\sigma^2,\zeta)}\rmQ^{-1}(\varepsilon).
\end{align}
\end{theorem}
The proof of Theorem \ref{secondorder} is provided in Section \ref{sec:proofsecondorder}. In the proof, we show that for any $\varepsilon\in(0,1)$,
\begin{align}
\log M_{\rm{sp}}^*(n,\varepsilon,\sigma^2,D)=
\log M_{\rm{iid}}^*(n,\varepsilon,\sigma^2,D)
&=\frac{n}{2}\log\frac{\sigma^2}{D}+\sqrt{n\rmV(\sigma^2,\zeta)}\rmQ^{-1}(\varepsilon)+O(\log n)\label{spheresecond}.
\end{align}
A few remarks are in order. 

First, in contrast to Scarlett, Tan, and Durisi~\cite{scarlett2017mismatch} where spherical and i.i.d.\ Gaussian codebooks achieve different second-order coding rates for the channel coding saddle-point problem (where the codebook is Gaussian and the channel is additive and its noise is non-Gaussian)~\cite{lapidoth1996}, we observe {\em no performance gap} in second-order asymptotics between two kinds of Gaussian codebooks in the rate-distortion saddle-point counterpart. We provide some intuition why this is so. In the rate-distortion problem, it is sufficient to use roughly $ \exp(\frac{n}{2}\log\frac{\sigma^2}{D})$ codewords to cover the set of typical source sequences (using either the spherical or i.i.d.\ Gaussian codebook) with the probability of failure in covering the typical sequences decaying super-exponentially. Thus, the probability that a typical source sequence remains uncovered is  vanishingly  small. Therefore, the dominant error event is the atypicality of the source sequence regardless which codebook ensemble is used.

Second, for a GMS (i.e., $X\sim\calN(0,\sigma^2)$), the dispersion is known to be $\rmV(\sigma^2,\zeta)=\frac{1}{2}$ independent of the distortion level or the source variance \cite{kostina2012fixed,ingber2011dispersion}. Hence, when specialized to GMSes, our results are consistent with existing results, namely that both spherical and i.i.d.\ Gaussian  codebooks achieve the optimal second-order coding rate for the rate-distortion problem~\cite[Theorems~13 and~40]{kostina2012fixed}. Furthermore, our results strengthen those of Lapidoth~\cite[Theorem~3]{lapidoth1997} in the sense that, using i.i.d.\  Gaussian codebooks, it is true that for an arbitrary memoryless source satisfying \eqref{sourceconstraint} and for any $\varepsilon\in(0,1)$, the first-order result in~\eqref{iidfotrue} holds. 

Finally, we mention in passing that our proof strategy differs significantly from Lapidoth's~\cite{lapidoth1997} who, for the direct and ensemble converse parts, invoked a theorem due to Wyner~\cite{Wyner67} concerning packings and coverings of $n$-spheres. The analysis used to prove this theorem and the subsequent ones naturally require more refined estimates on various probabilities.
\subsection{Moderate Deviation Asymptotics }
\begin{theorem}
\label{mdd}
Consider an arbitrary memoryless source $X$ satisfying \eqref{sourceconstraint} and $\Lambda_{X^2}(\theta)$ is finite for some positive number $\theta$. If $\rmV(\sigma^2,\zeta)$ is positive,\begin{align}
\nu^*_{\rm{sp}}=
\nu^*_{\rm{iid}}=
\frac{1}{2\rmV(\sigma^2,\zeta)}.
\end{align}
\end{theorem}
The proof of Theorem \ref{mdd} is provided  in Section \ref{sec:proofmdd}. 

We remark that $\xi_n=n^{-t}$ for any $t\in(0,1/2)$ satisfies  \eqref{req:xin}.  Notice though that  the second condition in~\eqref{req:xin} is more stringent compared to the more common moderate deviations condition in~\cite{altugwagner2014,polyanskiy2010channel}, namely $\sqrt{n} \xi_n\to\infty$. We believe that~\eqref{req:xin} is not fundamental and may be relaxed to $\sqrt{n} \xi_n\to\infty$ but a more refined analysis is required.

\subsection{Large Deviation Asymptotics }
We present several definitions before stating our main result. Given $s\in\bbR$ and any non-negative number $z$, define
\begin{align}
R_{\rm{sp}}(z)&:=-\frac{1}{2}\log\bigg( 1-\frac{(z+\sigma^2-2D)^2}{4z(\sigma^2-D)}\bigg)\label{def:rspa2},\\
R_{\rm{iid}}(s,z)&:=\frac{1}{2}\log(1+2s)+\frac{sz}{(1+2s)(\sigma^2-D)}-\frac{sD}{\sigma^2-D}\label{def:riidsapha},\quad\mbox{and}\\
s^*(z)&:=\max\bigg\{0,\frac{\sigma^2-3D+\sqrt{(\sigma^2-D)^2+4zD}}{4D}\bigg\}\label{def:sstar}.
\end{align}
We remark that $R_{\rm{sp}}(z)$ is the rate of the exponential decay of the non-excess-distortion probability for any source sequence $x^n$ whose power is $z=\frac{1}{n}{\|x^n\|^2}$ when its reproduction sequence is generated according to $f_{Y^n}^{\mathrm{sp}}$ in~\eqref{spherecodebook}, i.e.,
\begin{align}
\lim_{n\to\infty}-\frac{1}{n}\log \Pr\{d(x^n,Y^n)\leq D\}=R_{\rm{sp}}(z),\quad\mbox{where}\quad Y^n\sim f_{Y^n}^{\mathrm{sp}}.
\end{align} 
Similarly, $R_{\rm{iid}}(s^*(z),z)$ is the exponent of the non-excess-distortion probability for any source sequence $x^n$ with power $z$ when $Y^n\sim  f_{Y^n}^{\mathrm{iid}}$.

Recall that $\Lambda_{X^2}^*(t)$ is the Fenchel-Legendre transform of the cumulant generating function of $X^2$ (cf. Section \ref{sec:notation}), which is also known as the {\em large deviations rate function}~\cite{dembo2009large} of $X^2$. For brevity in presentation of the following theorem, let
\begin{align}
r_2&:=\sqrt{\sigma^2-D}+\sqrt{D}\label{def:r2}.
\end{align}

\begin{theorem}
\label{ldresult}
The following results on the ensemble excess-distortion exponents hold.
\begin{itemize}
\item  If $R<\frac{1}{2}\log\frac{\sigma^2}{D}$, then for both spherical and i.i.d.\ Gaussian codebooks,
\begin{align}
E^*_{\rm{sp}}(R)=E^*_{\rm{iid}}(R)=0.
\end{align}
\item If $R\geq \frac{1}{2}\log\frac{\sigma^2}{D}$, 
\begin{itemize}
\item For the spherical codebook, the ensemble excess-distortion exponent satisfies 
\begin{align}
E^*_{\rm{sp}}(R)=\Lambda_{X^2}^*(\alpha)\label{boundsldsphere},
\end{align}
where $\alpha\in[\sigma^2,r_2^2)$ is implicitly determined by $R$ through the equation
\begin{align}
R=R_{\rm{sp}}(\alpha)\label{ratesphereee}.
\end{align}

\item For the i.i.d.\ Gaussian codebook, the ensemble excess-distortion exponent is
\begin{align}
E^*_{\rm{iid}}(R)=\Lambda_{X^2}^*(\alpha)\label{ree:iid},
\end{align}
where $\alpha\geq \sigma^2$ is implicitly  determined by $R$ through the equation
\begin{align}
R&=R_{\rm{iid}}(s^*(\alpha),\alpha)\label{rateiidee}.
\end{align}
\end{itemize}
\end{itemize}
\end{theorem}
The proof of Theorem \ref{ldresult} is provided in Section \ref{sec:proofld}. Several remarks are in order.

First, as can be gleaned in the proof of Theorem \ref{ldresult} (cf. Lemma \ref{property}), $R_{\rm{sp}}(z)$ and $R_{\rm{iid}}(s^*(z),z)$ are both increasing functions of $z$ if $z\geq \sigma^2$ and $R_{\rm{sp}}(\sigma^2)=R_{\rm{iid}}(s^*(\sigma^2),\sigma^2)=\frac{1}{2}\log\frac{\sigma^2}{D}$. Furthermore, $\Lambda_{X^2}^*(t)>0$ for $t>\sigma^2$ and $\Lambda_{X^2}^*(t)=0$ otherwise. Combining these two facts, we can conclude that $E_{\rm{sp}}^*(R)$ and $E_{\rm{iid}}^*(R)$ are both positive for rates $R>\frac{1}{2}\log\frac{\sigma^2}{D}$. That $E_{\rm{sp}}^*(R)>0$ for $R>\frac{1}{2}\log\frac{\sigma^2}{D}$  recovers the achievability part of \cite[Theorem~3]{lapidoth1997} without recourse to Wyner's theorem~\cite{Wyner67}.

Second, for the i.i.d.\ Gaussian  codebook, if we consider that each codeword is generated according to \eqref{iidcodebook} with $\sigma^2$ replaced by $\alpha\in\bbR_+$, then the right hand side of \eqref{rateiidee} is replaced by $\frac{1}{2}\log\frac{\alpha}{D}$. Under this scenario, by particularizing the result to a GMS, we can recover the achievability result of Ihara and Kubo~\cite{ihara2000error}.\footnote{Note that we use $\alpha$ while Ihara and Kubo~\cite{ihara2000error} use $\alpha^2$ to mean the same quantity.} 

In the following lemma, we compare between  $E_{\rm{iid}}^*(R)$ and $E_{\rm{sp}}^*(R)$. 
\begin{lemma}
\label{comld}
For any rate $R>\frac{1}{2}\log\frac{\sigma^2}{D}$, we have
\begin{align}
E_{\rm{iid}}^*(R)>E_{\rm{sp}}^*(R).
\end{align}
\end{lemma}
The proof of Lemma \ref{comld} is provided in Appendix \ref{proofcomld}. Intuitively, Lemma \ref{comld} follows due to a subtle difference between  the spherical and i.i.d.\ Gaussian codebooks in the large deviations regime. When we use a spherical codebook, the non-excess-distortion event occurs with probability exactly zero for atypical source sequences with too small or too large powers. However, when we use an i.i.d.\ Gaussian codebook, the non-excess-distortion event occurs with some non-zero probability even for atypical source sequences. Note that the probability of this set of atypical sequences is exponentially small and thus does not lead to different performances of the two codebooks in the second-order and moderate deviations regimes but it does so in the large deviations regime.

\begin{figure}[t]
\centering
 \subfigure[Discrete]{\includegraphics[width=9cm]{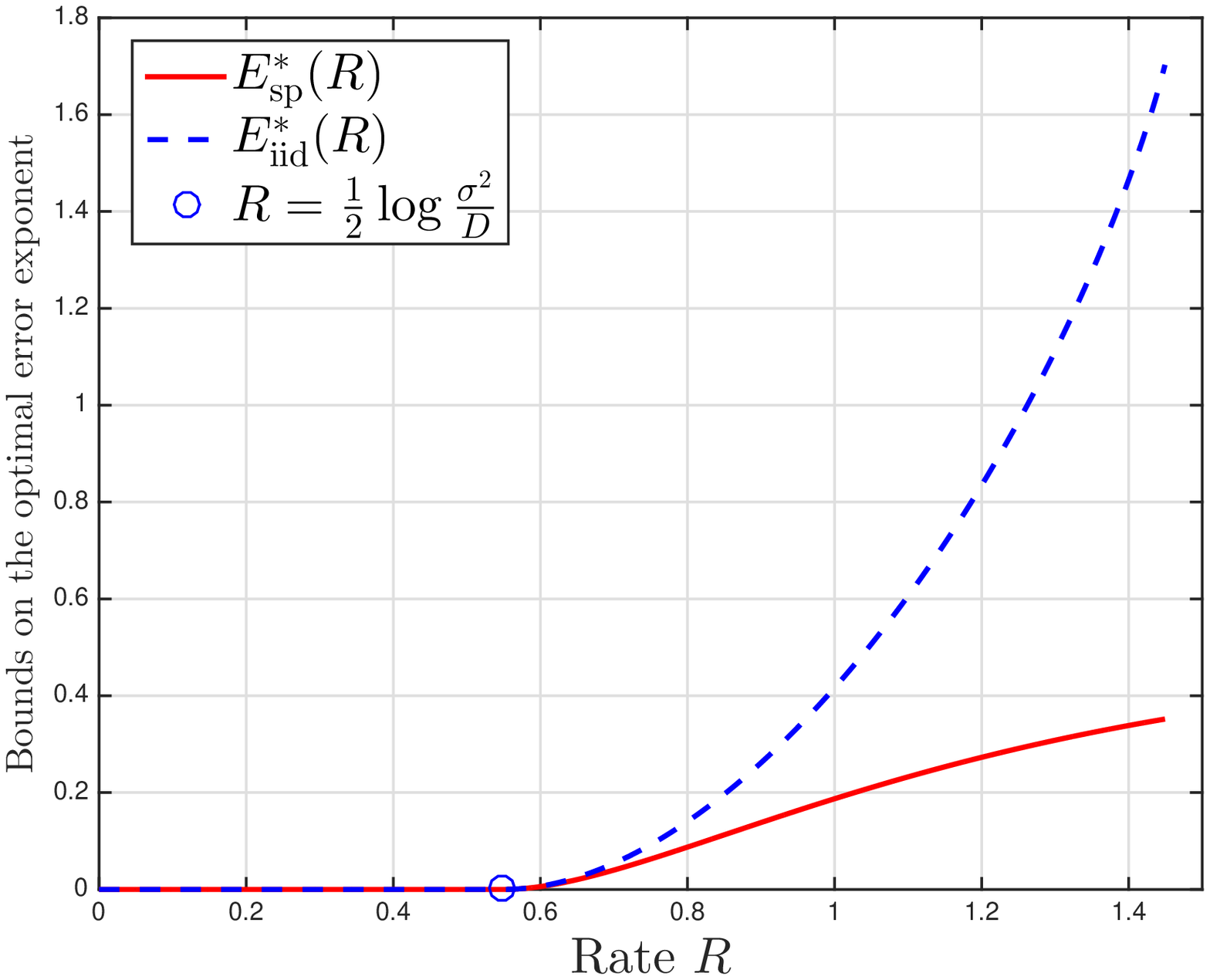}}
 \subfigure[Rayleigh]{\includegraphics[width=9cm]{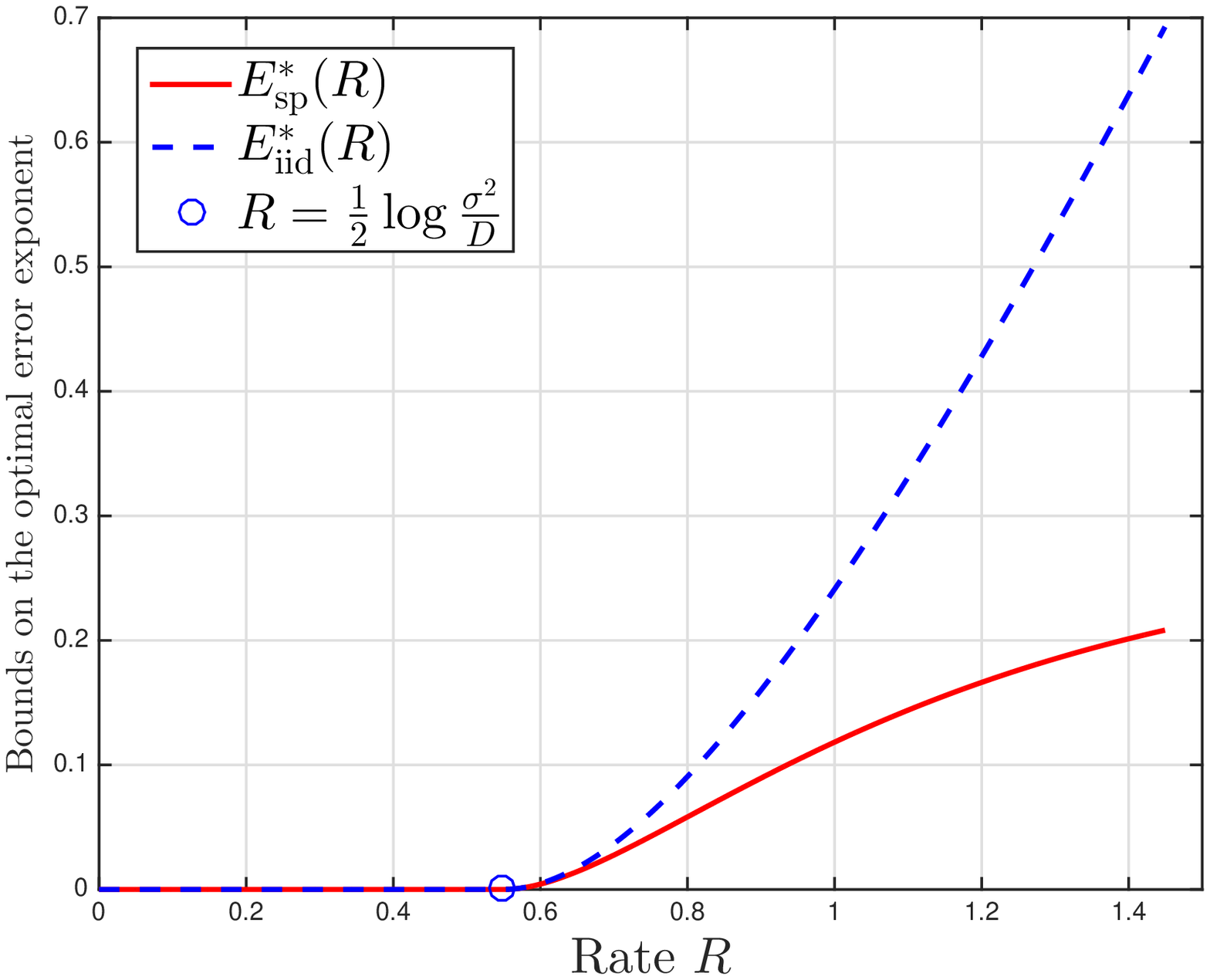}}
 \caption{Excess-distortion exponents $E_{\rm{sp}}^*(R)$ and $E_{\rm{iid}}(R)$ for a memoryless source distributed according to a discrete and a Rayleigh distribution. The discrete distribution is ternary with support $\calX=\{a,2a,3a\}$ where $a^2=0.3\sigma^2$ and its probability mass function is $P_X= [ 1/2,1/3,1/6 ]$  so that $\bbE[X^2]=\sigma^2$. The Rayleigh distribution has   scale parameter $\sigma^2/2$  so that $\bbE[X^2]=\sigma^2$.  Note that $\alpha$ is determined through $R$ by~\eqref{ratesphereee} for the spherical codebook and by~\eqref{rateiidee} for the i.i.d.\ Gaussian codebook. }
 \label{boundld}
\end{figure}

To illustrate the result in Lemma \ref{comld}, we plot ensemble excess-distortion exponents of spherical and i.i.d.\ Gaussian codebooks for a discrete and a Rayleigh distribution in Figure \ref{boundld}. From Figure \ref{boundld}, we observe that for both numerical examples, the i.i.d.\ Gaussian codebook has a strictly larger excess-distortion exponent than the spherical codebook for any rate $R>\frac{1}{2}\log\frac{\sigma^2}{D}$.

\section{Proof of Second-Order Asymptotics (Theorem \ref{secondorder})}
\label{sec:proofsecondorder}
\subsection{Preliminaries for the Spherical Codebook}
\label{sec:presphere}
In this subsection, we present some definitions and preliminary results for spherical codebooks. For simplicity, let the variance or power of $Y$ be $P_Y:=\sigma^2-D$. Furthermore, for any $\varepsilon\in(0,1)$, let 
\begin{align}
\rmV&:=\mathrm{Var}[X^2]=\zeta-\sigma^4\label{def:rmV},\\
a_n&:=\sqrt{\rmV\frac{\log n}{n}},\label{def:an}\\
b_n&:=\sqrt{\frac{\rmV}{n}}\rmQ^{-1}(\varepsilon)\label{def:bn},
\end{align}
where the second equality in \eqref{def:rmV} follows from the definition in \eqref{sourceconstraint}. Note that for any $x^n$, $\Pr\{d(x^n,Y^n)\leq D\}$ depends on $x^n$ only through its norm $\|x^n\|$. For any $x^n$ such that $\frac{1}{n}  {\|x^n\|^2}=z>0$, let 
\begin{align}
\Psi(n,z)&:=\Pr\{d(x^n,Y^n)\leq D\}\\
&=\Pr\{\|x^n-Y^n\|^2\leq nD\}\\
&=\Pr\{\|x^n\|^2+\|Y^n\|^2-2\langle x^n,Y^n \rangle\leq nD\}\label{ensemble}\\
&=\Pr\{nz+nP_Y-2\langle x^n,Y^n\rangle \leq nD\}\\
&=\Pr\{2\langle x^n,Y^n\rangle \geq n(z+P_Y-D)\}\\
&=\Pr\bigg\{Y_1\geq \frac{\sqrt{n}(z+P_Y-D)}{2\sqrt{z}}\bigg\}\label{y1symm},
\end{align}
where $Y_1$ is the first element of sequence $Y^n = (Y_1,\ldots, Y_n)$ and \eqref{y1symm} follows because $Y^n$ is spherically symmetric so we may take $x^n =( \sqrt{nz},0,\ldots, 0)$ (cf.~\cite{scarlett2017mismatch}).

Let $Z:=\frac{1}{n}{\|X^n\|^2}$ be the random variable representing the average power of the source  $X^n$. Furthermore, let $f_Z$ be the corresponding probability distribution function (pdf) of $Z$. Recall that $r_2=\sqrt{\sigma^2-D}+\sqrt{D}=\sqrt{P_Y}+\sqrt{D}$ (cf. \eqref{def:r2}) and let
\begin{align}
r_1&:=\sqrt{P_Y}-\sqrt{D}\label{def:r1}.
\end{align}

Kostina and Verd\'u~\cite[Theorem 37]{kostina2012fixed} showed that for any $z$ such that $\sqrt{z}<r_1$ or $\sqrt{z}>r_2$,
\begin{align}
\Psi(n,z)=0,\label{psi=1foratp}
\end{align}
and otherwise
\begin{align}
\Psi(n,z)\geq \frac{\Gamma(\frac{n+2}{2})}{\sqrt{\pi} n\Gamma(\frac{n+1}{2})}\bigg(1-\frac{(z+P_Y-D)^2}{4zP_Y}\bigg)^{\frac{n-1}{2}} = :\underline{g}(n,z),\label{def:undergnz}
\end{align}
where $\Gamma(\cdot)$ is the Gamma function. Hence, from \eqref{peforboth} and \eqref{psi=1foratp}, we conclude that the excess-distortion probability for the spherical codebook is
\begin{align}
\rmP_{\rme,n}(M)
&=\Pr\{Z<(\max\{0,r_1\})^2\}+\Pr\{Z>r_2^2\}+\int_{(\max\{0,r_1\})^2}^{r_2^2}(1-\Psi(n,z))^M f_Z(z)\, \rmd z\label{pe4sphere}.
\end{align}

\subsection{Achievability Proof for the spherical Codebook}
\label{sec:sphereach}

Using the definition of $\underline{g}(\cdot)$ in \eqref{def:undergnz}, we conclude that $\underline{g}(n,z)$ is a decreasing function of $z$ if $z\geq |P_Y-D|$. Invoking the definitions of $b_n$ in \eqref{def:bn}, $r_1$ in \eqref{def:r1} and $r_2$ in \eqref{def:r2}, we conclude that $r_1^2\leq |P_Y-D|$ and $r_2^2\geq \sigma^2+b_n$ for $n$ large enough.
Thus, combining \eqref{def:undergnz}, \eqref{pe4sphere} and noting that $\Psi(n,z)\geq 0$, for sufficiently large $n$, we can upper bound the excess-distortion probability as follows:
\begin{align}
\rmP_{\rme,n}(M)
&\leq \Pr\{Z<|P_Y-D|\}+\int_{|P_Y-D|}^{\sigma^2+b_n}(1-\underline{g}(n,z))^M f_Z(z)\, \rmd z+\Pr\{Z>\sigma^2+b_n\}\label{r2c4}\\
&\leq \Pr\{Z<|P_Y-D|\}+\int_{|P_Y-D|}^{\sigma^2+b_n}\exp\{-M\underline{g}(n,z)\}f_Z(z)\, \rmd z+\Pr\{Z>\sigma^2+b_n\}\label{upplog1-x},\\
&\leq \Pr\{Z<|P_Y-D|\}+\exp\{-M\underline{g}(n,\sigma^2+b_n)\}+\Pr\{Z>\sigma^2+b_n\}\label{usenondecreaseug}
\end{align}
where \eqref{upplog1-x} follows since $(1-a)^M\leq \exp\{-Ma\}$ for any $a\in[0,1)$; and \eqref{usenondecreaseug} follows since $\underline{g}(n,z)$ is decreasing in $z$ for $z\geq |P_Y-D|$. Let the third central moment of $X^2$ be defined as
\begin{align}
T:=\bbE\big[|X^2-\sigma^2|^3\big]\label{def:T}.
\end{align}
Using the definitions of $\rmV$ in \eqref{def:rmV}, $T$ in \eqref{def:T} and the Berry-Esseen theorem, we conclude that
\begin{align}
\Pr\{Z<|P_Y-D|\}
&=\Pr\bigg\{\frac{1}{n}\sum_{i=1}^n X_i^2<|\sigma^2-2D|\bigg\}\\
&\leq \frac{6T}{\sqrt{n}\, \rmV^{3/2}}+\rmQ\bigg( \big(\sigma^2-|\sigma^2-2D|\big)\sqrt{\frac{n}{\rmV}}\bigg)\\
&\leq \frac{6T}{\sqrt{n}\, \rmV^{3/2}}+\exp\bigg\{-\frac{2n\big(\sigma^2-|\sigma^2-2D|\big)^2}{\rmV}\bigg\}\label{upprmq2}\\
&=O\bigg( \frac{1}{\sqrt{n}}\bigg),\label{achstep1}
\end{align}
where \eqref{upprmq2} follows since $\rmQ(a)\leq \exp\{-\frac{a^2}{2}\}$ while \eqref{achstep1} follows since $T$ (cf. \eqref{def:T}) is finite for sources satisfying \eqref{sourceconstraint} and $\sigma^2-|\sigma^2-2D|>0$ due to the fact that $\sigma^2>D$. Similarly, using the definition of $b_n$ in \eqref{def:bn} and the Berry-Esseen theorem, we have
\begin{align}
\Pr\{Z>\sigma^2+b_n\}
&=\Pr\bigg\{\frac{1}{n}\sum_{i=1}^nX_i^2>\sigma^2+b_n\bigg\}\\
&\leq \varepsilon+\frac{6T}{\sqrt{n}\, \rmV^{3/2}}\\
&\leq \varepsilon+O\bigg( \frac{1}{\sqrt{n}}\bigg)\label{achstep2}.
\end{align}

Choose $M$ such that 
\begin{align}
\log M
&=-\log \underline{g}(n,\sigma^2+b_n)+\log\bigg( \frac{1}{2}\log n\bigg)\label{hi:chosem}\\
&=n\bigg( \frac{1}{2}\log \frac{\sigma^2}{D}+\frac{b_n}{2\sigma^2}+O\bigg( \frac{\log n}{n}\bigg)\bigg)\label{tayloragain}\\
&=\frac{n}{2}\log \frac{\sigma^2}{D}+\sqrt{n\rmV(\sigma^2,\zeta)}\rmQ^{-1}(\varepsilon)+O(\log n)\label{usebnha},
\end{align}
where \eqref{tayloragain} follows from the Taylor expansion of $\underline{g}(n,\sigma^2+b_n)$ (cf. \eqref{def:undergnz}) and noting that $ {\Gamma(\frac{n+2}{2})}/{\Gamma(\frac{n+1}{2})}=\Theta(\sqrt{n})$; and \eqref{usebnha} follows from the definition of $b_n$ (cf. \eqref{def:bn}) and $\rmV(\sigma^2,D)$ (cf. \eqref{def:rmvsphere}). Thus, with the choice of $M$ in \eqref{hi:chosem}, we conclude that
\begin{align}
\exp\{-M\underline{g}(n,\sigma^2+b_n)\}&=\frac{1}{\sqrt{n}}\label{achstep3}.
\end{align}
Hence, combining \eqref{usenondecreaseug}, \eqref{achstep1}, \eqref{achstep2}, \eqref{usebnha} and \eqref{achstep3}, we have shown  that
\begin{align}
\log M_{\rm{sp}}^*(n,\varepsilon,\sigma^2,D)
&\geq \frac{n}{2}\log \frac{\sigma^2}{D}+\sqrt{n\rmV(\sigma^2,\zeta)}\rmQ^{-1}(\varepsilon)+O(\log n).
\end{align}

\subsection{Ensemble Converse for the spherical Codebook}
\label{sec:spheresecconverse}
We now show that the result in \eqref{spheresecond} is ensemble tight. From Stam's paper \cite[Eq. (4)]{stam1982limit}, the distribution of $Y_1$ is 
\begin{align}
f_{Y_1}(y)=\frac{1}{\sqrt{\pi n P_Y}}\frac{\Gamma(\frac{n}{2})}{\Gamma(\frac{n-1}{2})}\bigg( 1-\frac{y^2}{nP_Y}\bigg)^{\frac{n-3}{2}}1\{y^2\leq nP_Y\}\label{pdfy1}.
\end{align}
Recall the definitions of $a_n$ in \eqref{def:an} and $b_n$ in \eqref{def:bn}. Define the sets 
\begin{align}
\calP&:=\{r\in\bbR:b_n<r-\sigma^2\leq a_n\}\label{def:calp},\\
\calQ&:=\{r\in\bbR:r+P_Y-D\geq 0\}\label{def:calq}.
\end{align}
Then, for any $z\in\calP\cap\calQ$ satisfying $\frac{\sqrt{n}(z+P_Y-D)}{2\sqrt{z}}\leq \sqrt{nP_Y}$, using the definition of $\Psi(\cdot)$ in \eqref{y1symm}, we obtain that
\begin{align}
\Psi(n,z)
&=\Pr\bigg\{Y_1\geq \frac{\sqrt{n}(z+P_Y-D)}{2\sqrt{z}}\bigg\}\\
&= \int_{\frac{\sqrt{n}(z+P_Y-D)}{2\sqrt{z}}}^{\sqrt{nP_Y}}\frac{1}{\sqrt{\pi n P_Y}}\frac{\Gamma(\frac{n}{2})}{\Gamma(\frac{n-1}{2})}\bigg(1-\frac{y^2}{nP_Y}\bigg)^{\frac{n-3}{2}}\rmd y\label{usepdfy1}\\
&\leq \int_{\frac{\sqrt{n}(z+P_Y-D)}{2\sqrt{z}}}^{\sqrt{nP_Y}}\frac{1}{\sqrt{\pi n P_Y}}\frac{\Gamma(\frac{n}{2})}{\Gamma(\frac{n-1}{2})}\bigg(1-\frac{(z+P_Y-D)^2}{4zP_Y}\bigg)^{\frac{n-3}{2}}\rmd y\label{nonincreasing}\\
&\leq \frac{1}{\sqrt{\pi}}\frac{\Gamma(\frac{n}{2})}{\Gamma(\frac{n-1}{2})}\bigg(1-\frac{(z+P_Y-D)^2}{4zP_Y}\bigg)^{\frac{n-3}{2}}\label{enlargeinterval}\\
&=\frac{1}{\sqrt{\pi}}\frac{\Gamma(\frac{n}{2})}{\Gamma(\frac{n-1}{2})}\exp\bigg\{\frac{n-3}{2}\log \bigg( 1-\frac{(z+P_Y-D)^2}{4zP_Y}\bigg)\bigg\}=:\overline{g}(n,z)\label{def:overlineg},
\end{align}
where \eqref{usepdfy1} follows from the definition in \eqref{pdfy1} and the condition that $z\in\calQ$ (cf. \eqref{def:calq}) which implies $\frac{\sqrt{n}(z+P_Y-D)}{2\sqrt{z}}\geq 0>-\sqrt{nP_Y}$; \eqref{nonincreasing} follows since $(1-\frac{y^2}{nP_Y})$ is decreasing in $y$ for positive $y$; and \eqref{enlargeinterval} follows by enlarging the integration region (recall that $\frac{\sqrt{n}(z+P_Y-D)}{2\sqrt{z}}\geq 0$). Note that $\overline{g}(n,z)$ is decreasing in $z$ for $z\geq |P_Y-D|$ and $\overline{g}(n,z)\geq 0$ for all $z\in\calP$. Hence, for any $z\in\calP\cap\calQ$ such that $\frac{\sqrt{n}(z+P_Y-D)}{2\sqrt{z}}>\sqrt{nP_Y}$, we still have $\overline{g}(n,z)\geq \Psi(n,z)$. 

Recall that $Z=\frac{1}{n} {\|X^n\|^2}$ and $f_Z$ is the corresponding pdf of $Z$. Thus, according to \eqref{peforboth}, for $n$ sufficiently large, we have 
\begin{align}
\rmP_{\rme,n}(M)
&=\bbE_{X^n}[(1-\Pr\{d(X^n,Y^n)\leq D|X^n\})^M]\label{fromhere}\\
&=\int_0^\infty (1-\Psi(n,z))^M f_Z(z)\, \rmd z\label{usepsidef}\\
&\geq \int_0^\infty (1-\overline{g}(n,z))^M 1\{z\in\calP\cap\calQ\}f_Z(z)\, \rmd z\label{usedefg}\\
&\geq \int_{z\in\calP\cap\calQ} (1-\overline{g}(n,\sigma^2+b_n))^M f_Z(z)\, \rmd z\label{usenonincreaseoverg}\\
&\geq \int_{z\in\calP\cap\calQ} \exp\bigg\{-M\frac{\overline{g}(n,\sigma^2+b_n)}{1-\overline{g}(n,\sigma^2+b_n)}\bigg\}f_Z(z)\, \rmd z\label{lblog}\\
&\geq \int_{z\in\calP\cap\calQ}\exp\bigg\{-M\frac{\overline{g}(n,\sigma^2+b_n)}{1-\overline{g}(n,\sigma^2+b_n)}\bigg\}
1\bigg\{M\frac{\overline{g}(n,\sigma^2+b_n)}{1-\overline{g}(n,\sigma^2+b_n)}\leq \frac{1}{\sqrt{n}}\bigg\} \, f_Z(z)\rmd z\\
&\geq \bigg( 1-\frac{1}{\sqrt{n}}\bigg)\int_{z\in\calP\cap\calQ}
1\bigg\{M\frac{\overline{g}(n,\sigma^2+b_n)}{1-\overline{g}(n,\sigma^2+b_n)}\leq \frac{1}{\sqrt{n}}\bigg\} \, f_Z(z)\rmd z\label{addconstraint}\\
&=\bigg( 1-\frac{1}{\sqrt{n}}\bigg)\Pr\bigg\{Z\in\calP\cap\calQ,M\leq \frac{1-\overline{g}(n,\sigma^2+b_n)}{\overline{g}(n,\sigma^2+b_n)}\frac{1}{\sqrt{n}}\bigg\}\\
&=\bigg( 1-\frac{1}{\sqrt{n}}\bigg)\Pr\bigg\{Z\in\calP\cap\calQ,\log M\leq \log(1-\overline{g}(n,\sigma^2+b_n))
-\log \overline{g}(n,\sigma^2+b_n)-\frac{1}{2}\log n\bigg\}\\
&\geq \bigg( 1-\frac{1}{\sqrt{n}}\bigg)\Pr\bigg\{Z\in\calP\cap\calQ,\log M\leq -\log 2
-\log \overline{g}(n,\sigma^2+b_n)-\frac{1}{2}\log n\}\bigg\}\label{nlarge},
\end{align}
where \eqref{usepsidef} follows from the definition of $\Psi(n,z)$ in \eqref{y1symm}; \eqref{usedefg} follows by restricting $z\in\calP\cap\calQ$ and using the definition of $\overline{g}(\cdot)$ in \eqref{def:overlineg}; \eqref{usenonincreaseoverg} follows since $\overline{g}(n,z)$ is decreasing in $z$ for $z\in\calP\cap\calQ$; \eqref{lblog} follows since $(1-a)^M\geq \exp\{-M\frac{a}{1-a}\}$ for any $a\in[0,1)$; \eqref{addconstraint} follows since $M\frac{\overline{g}(n,z)}{1-\overline{g}(n,z)}\leq \frac{1}{\sqrt{n}}$, $\exp\{-a\}$ is decreasing in $a$, and $\exp\{-a\}\geq 1-a$ for $a\geq 0$; and \eqref{nlarge} follows since $\overline{g}(n,z)\leq \frac{1}{2}$ for $n$ large enough if $z>\sigma^2$.

Combining \eqref{def:overlineg}, \eqref{nlarge} and applying a Taylor expansion of $\overline{g}(n,\sigma^2+b_n)$ similarly to \eqref{tayloragain}, we conclude that for any $(n,M)$-code such that
\begin{align}
\log M
&\leq -\log 2-\frac{1}{2}\log n-\log \overline{g}(n,\sigma^2+b_n)\\
&=n\bigg( \frac{1}{2}\log\frac{\sigma^2}{D}+\frac{b_n}{2\sigma^2}+O\bigg( \frac{\log n}{n}\bigg)\bigg)\label{usezinp},
\end{align}
we have
\begin{align}
\rmP_{\rme,n}(M)
&\geq \bigg(1-\frac{1}{\sqrt{n}}\bigg)\Pr\{Z\in\calP\cap\calQ\}\label{conversestep3}.
\end{align}
The following lemma is essential to complete  the converse proof.
\begin{lemma}
\label{concentrate}
Consider any source $X$ such that \eqref{sourceconstraint} are satisfied and $\sigma^2<\infty$. Then, we have 
\begin{align}
\Pr\{Z\in\calP\cap\calQ\}
\geq \varepsilon+O\bigg( \frac{1}{\sqrt{n}}\bigg).
\end{align}
\end{lemma}
The proof of Lemma \ref{concentrate} is deferred to Appendix \ref{proof:concentrate}.

Using the definition of $\rmV(\sigma^2,\zeta)$ in \eqref{def:rmvsphere}, the definition of $b_n$ in \eqref{def:bn}, the bounds in \eqref{usezinp}, \eqref{conversestep3}, and Lemma \ref{concentrate}, we conclude that 
\begin{align}
\log M_{\rm{sp}}^*(n,\varepsilon,\sigma^2,D)
&\leq \frac{n}{2}\log \frac{\sigma^2}{D}+\sqrt{n\rmV(\sigma^2,\zeta)}\rmQ^{-1}(\varepsilon)+O(\log n).
\end{align}


\subsection{Preliminaries for the  I.I.D. Gaussian Codebook}
\label{sec:preiid}
Now we consider the i.i.d.\ Gaussian codebook (cf.~\eqref{iidcodebook}). Note that $\Pr\{d(x^n,Y^n)\leq D\}$ depends on $x^n$ only through its norm $\|x^n\|$ (cf. \cite{ihara2000error}). Given any sequence $x^n$ such that $\frac{1}{n} {\|x^n\|^2} =z$, define
\begin{align}
\Upsilon(n,z)
&:=\Pr\{d(x^n,Y^n)\leq D\}\label{def:upsilonnz}.
\end{align}
From \eqref{iidcodebook}, we obtain that
\begin{align}
f_{Y^n}^{\rm{iid}}(y^n)&=\frac{1}{(2\pi(\sigma^2-D))^{n/2}}\exp\bigg\{-\frac{\|y^n\|^2}{2(\sigma^2-D)}\bigg\}.
\end{align}
Since $f_{Y^n}^{\rm{iid}}(y^n)$ is decreasing in $\|y^n\|$, we conclude that $\Upsilon(n,z)$ is a decreasing function of $z$ (cf. \cite{ihara2000error}). Using the definition of $\Upsilon(\cdot)$ in \eqref{def:upsilonnz}, we have
\begin{align}
\Upsilon(n,z)
&=\Pr\{\|x^n-Y^n\|^2\leq nD\}\\
&=\Pr\bigg\{\sum_{i=1}^n(Y_i-\sqrt{z})^2\leq nD\bigg\}\label{powerinvariant}\\
&=\Pr\bigg\{-\frac{1}{nP_Y}\sum_{i=1}^n(Y_i-\sqrt{z})^2\geq -\frac{D}{P_Y}\bigg\}.
\end{align}
where \eqref{powerinvariant} follows since the probability depends on $x^n$ only through its power and thus we can choose $x^n$ such that  $x_i=\sqrt{z}$ for all $i\in[1:n]$ (cf.~\cite[Eq. (94)]{scarlett2017mismatch}). For the i.i.d.\ Gaussian codebook, each $Y_i\sim\calN(0,P_Y)$ and hence $\frac{1}{P_Y} {(Y_i-\sqrt{z})^2}$ is distributed according to a non-central $\chi^2$ distribution with one degree of freedom.

Given $z$ and $s$, let
\begin{align}
\kappa(s,z)&:=\frac{(P_Y(1+2s)+2z)^2}{P_Y(1+2s)^3}\label{def:lambda2s}
\end{align}
Using the result of \cite[Section 2.2.12]{tanizaki2004computational} concerning the cumulant generating function of a non-central $\chi^2$ distribution, the definition of $R_{\rm{iid}}(\cdot)$ in \eqref{def:riidsapha}, the definition of $s^*(\cdot)$ in \eqref{def:sstar}, and the Bahadur-Ranga Rao (strong large deviations) theorem for non-lattice random variables~\cite[Theorem~3.7.4]{dembo2009large},  we obtain  
\begin{align}
\Upsilon(n,z)
&\sim\frac{\exp\{-nR_{\rm{iid}}(s^*(z),z)\}}{s^*(z)\sqrt{\kappa(s^*(z),z)2\pi n}} ,\quad n\to\infty\label{iidach1}.
\end{align}

\subsection{Achievability Proof for the I.I.D. Gaussian Codebook}
\label{sec:iidach}
According to \eqref{peforboth}, the excess-distortion probability under the i.i.d.\ Gaussian codebook can be upper bounded as follows:
\begin{align}
\rmP_{\rme,n}(M)
&=\bbE\Big[(1-\Pr\{d(X^n,Y^n)\leq D\,\big|\,X^n\})^M \Big]\label{ce1}\\
&=\int_0^\infty (1-\Upsilon(n,z))^Mf_Z(z)\, \rmd z\\
&\leq \int_0^{\sigma^2-a_n}f_Z(z)\, \rmd z+\int_{\sigma^2-a_n}^{\sigma^2+b_n} (1-\Upsilon(n,z))^Mf_Z(z)\, \rmd z+\int_{\sigma^2+b_n}^\infty f_Z(z)\, \rmd z\label{uppby1}\\
&\leq \int_{\sigma^2-a_n}^{\sigma^2+b_n} \exp\{-M\Upsilon(n,z)\}f_Z(z)\, \rmd z+\Pr\{Z<\sigma^2-a_n\}+\Pr\{Z>\sigma^2+b_n\}\label{useineq2}\\
&\leq \exp\{-M\Upsilon(n,\sigma^2+b_n) \}+\Pr\{Z<\sigma^2-a_n\}+\Pr\{Z>\sigma^2+b_n\}\label{iidach2},
\end{align}
where \eqref{uppby1} follows since $\Upsilon(n,z)\geq 0$; \eqref{useineq2} follows since $(1-a)^M\leq \exp\{-Ma\}$; and \eqref{iidach2} follows since $\Upsilon(n,z)$ is decreasing in $z$ and $\Pr\{\sigma^2-a_n\leq Z\leq \sigma^2+b_n\}\leq 1$.

Using the definitions of $R_{\rm{iid}}(\cdot)$ in \eqref{def:riidsapha} and $s^*(\cdot)$ in \eqref{def:sstar}, we have
\begin{align}
\nn&R_{\rm{iid}}(s^*(\sigma^2+b_n),\sigma^2+b_n)\\*
\nn&=\frac{1}{2}\log \frac{P_Y+\sqrt{P_Y^2+4(\sigma^2+b_n)D}}{2D}+\frac{z(P_Y-2D+\sqrt{P_Y^2+4(\sigma^2+b_n)D})}{2P_Y(P_Y+\sqrt{P_Y^2+4(\sigma^2+b_n)D})}\\*
&\qquad-\frac{P_Y-2D+\sqrt{P_Y^2+4(\sigma^2+b_n)D}}{4P_Y}\label{concavebys*}\\
&=\frac{1}{2}\log\frac{\sigma^2}{D}+\frac{b_n}{2\sigma^2}+O(b_n^2)\label{taylor3},\\
&=\frac{1}{2}\log\frac{\sigma^2}{D}+\sqrt{\frac{\rmV(\sigma^2,\zeta)}{n}}\rmQ^{-1}(\varepsilon)+O\bigg( \frac{1}{n}\bigg),\label{aftertaylot3}
\end{align}
where \eqref{taylor3} follows from a Taylor expansion at $z=\sigma^2$ and recalling that $P_Y=\sigma^2-D$; and \eqref{aftertaylot3} follows from the definitions of $\rmV(\sigma^2,\zeta)$ in \eqref{def:rmvsphere} and $b_n$ in \eqref{def:bn}.

Choose $M$ such that
\begin{align}
\log M\geq -\log \Upsilon(n,\sigma^2+b_n)+\log\bigg( \frac{1}{2}\log n\bigg).
\end{align}
Then, we have
\begin{align}
\exp\{-M\Upsilon(n,\sigma^2+b_n)\}
\leq \frac{1}{\sqrt{n}}\label{iidach3}.
\end{align}
Furthermore, using the result in \eqref{iidach1} and \eqref{aftertaylot3}, we obtain
\begin{align}
\log M
&\geq \frac{n}{2}\log\frac{\sigma^2}{d}+\sqrt{n\rmV(\sigma^2,\zeta)}\rmQ^{-1}(\varepsilon)+O(\log n)\label{iidachlogm4second}.
\end{align}
Similarly as the proof of Lemma \ref{concentrate}, using the Berry-Esseen theorem and the definition of $a_n$ in \eqref{def:an}, we obtain
\begin{align}
\Pr\{Z<\sigma^2-a_n\}
&=\Pr\bigg\{\frac{1}{n}\sum_{i=1}^n(X_i^2-\sigma^2)<\sqrt{\rmV\frac{\log n}{n}}\bigg\}\\
&\leq \rmQ(\sqrt{\log n})+\frac{6T}{\sqrt{n}\, \rmV^{3/2}}\\
&=O\bigg( \frac{1}{\sqrt{n}}\bigg)\label{iidach4}.
\end{align}

Hence, combining \eqref{achstep2}, \eqref{iidach2}, \eqref{iidach3}, \eqref{iidachlogm4second} and \eqref{iidach4}, we conclude that 
\begin{align}
\log M_{\rm{iid}}^*(n,\varepsilon,\sigma^2,D)
&\geq \frac{n}{2}\log \frac{\sigma^2}{D}+\sqrt{n\rmV(\sigma^2,\zeta)}\rmQ^{-1}(\varepsilon)+O(\log n).
\end{align}

\subsection{Ensemble Converse for the I.I.D. Gaussian Codebook}
\label{sec:iidconverse}

The ensemble converse proof for the i.i.d.\ Gaussian codebook is omitted since it is similar to the ensemble converse proof for the spherical codebook in Section \ref{sec:spheresecconverse} starting from \eqref{fromhere} except for the following two points: i) replace $\overline{g}(n,z)$ with $\Upsilon(n,z)$; ii) replace $\calP\cap\calQ$ with $\calP$.

\section{Proof of Moderate Deviation Asymptotics (Theorem \ref{mdd})}
\label{sec:proofmdd}
\subsection{Preliminaries}
We recall the following version of the Chernoff bound~\cite[Theorem B.4.1]{bouleau93}.
\begin{lemma}
\label{largediid} 
Given an i.i.d.\ sequence $X^n$, suppose that the cumulant generating function $\Lambda_{|X|}(\theta)$ is finite for some positive number $\theta$. Then for any $t>\bbE[X]$,
\begin{align}
\Pr\bigg\{\frac{1}{n}\sum_{i=1}^nX_i>t\bigg\}\leq \exp\{-n\Lambda_X^*(t)\}.
\end{align}       
and for any $t<\bbE[X]$,
\begin{align}
\Pr\bigg\{\frac{1}{n}\sum_{i=1}^nX_i<t\bigg\}\leq \exp\{-n\Lambda_X^*(t)\}.
\end{align}
In both cases, $\Lambda_X^*(t)>0$.
\end{lemma}
In other words, if  the threshold $t$ deviates from the mean by a constant, the probability in question decays exponentially fast.

\subsection{Achievability Proof for the spherical Codebook}
\label{sec:proofmddsphere}
Let 
\begin{align}
c_n:=2\sigma^2\xi_n\label{def:cn}.
\end{align}
The proof of the achievability follows from Section \ref{sec:sphereach} up till~\eqref{usenondecreaseug} with $c_n$ taking the role of $b_n$. Invoking Lemma \ref{largediid}, we conclude that under the conditions in Theorem \ref{mdd}, we have
\begin{align}
\Pr\{Z<|P_Y-D|\}
&=\Pr\bigg\{\frac{1}{n}\sum_{i=1}^nX_i^2<|P_Y-D|\bigg\}\\
&\leq \exp\{-nt_1\}\label{mdpart1}
\end{align}
for some $t_1>0$ since $|P_Y-D|=|\sigma^2-2D|<\sigma^2$ due to the fact that $\sigma^2>D$. Invoking the moderate deviations theorem~\cite[Theorem 3.7.1]{dembo2009large} and the definition of $\rmV$ in \eqref{def:rmV}, we conclude that
\begin{align}
\Pr\{Z>\sigma^2+c_n\}
&=\Pr\bigg\{\frac{1}{n}\sum_{i=1}^n (X_i^2-\sigma^2)>c_n\bigg\}\\
&=\exp\bigg\{-\frac{nc_n^2}{2\rmV}+o(nc_n^2)\bigg\}\\
&=\exp\bigg\{-\frac{n\xi_n^2}{2\rmV(\sigma^2,\zeta)}+o(n\xi_n^2)\bigg\},\label{mdpart2}
\end{align}
where \eqref{mdpart2} follows from the definitions of $\rmV(\sigma^2,\zeta)$ in \eqref{def:rmvsphere}, $\rmV$ in \eqref{def:rmV} and $c_n$ in \eqref{def:cn}.

Recall the definition of $\underline{g}(\cdot)$ in \eqref{def:undergnz}. Choose $M$ such that
\begin{align}
\log M
&=-\log \underline{g}(n,\sigma^2+c_n)+\log n\label{rateach0}\\
&=n\bigg( \frac{1}{2}\log \frac{\sigma^2}{D}+\xi_n+o(\xi_n)\bigg)\label{rateach},
\end{align}
where \eqref{rateach} follows from a Taylor expansion similar to \eqref{tayloragain}, the definition of $c_n$ in \eqref{def:cn}, and the conditions on $\xi_n$ in \eqref{req:xin}. With this choice of $M$, we have
\begin{align}
\exp\{-M\underline{g}(n,\sigma^2+c_n)\}= \exp(-n)\label{mdpart3}.
\end{align}
Combining the results in \eqref{usenondecreaseug}, \eqref{mdpart1}, \eqref{mdpart2}, \eqref{rateach} and \eqref{mdpart3},  we conclude that 
\begin{align}
\liminf_{n\to\infty}-\frac{1}{n\xi_n^2}\log\rmP_{\rme,n}(\exp(n (R(\sigma^2,D) +\xi_n) )
&\geq \frac{1}{2\rmV(\sigma^2,\zeta)}\label{achmddha}.
\end{align}

\subsection{Ensemble Converse Proof for  the spherical Codebook}\label{sec:proofmddsphere_conv}
Define the following set
\begin{align}
\calP'&:=\{r\in\bbR:\xi_n<r-\sigma^2\leq 2\xi_n\}\label{def:calp'n}. 
\end{align}

Following similar proof as in Section \ref{sec:spheresecconverse} up till~\eqref{conversestep3} with $(\calP',c_n)$  in place of $(\calP,b_n)$, we conclude that for any $(n,M)$-code such that
\begin{align}
\log M
&\leq -\log 2-\frac{1}{2}\log n-\log \overline{g}(n,\sigma^2+c_n),\\
&=n\bigg( \frac{1}{2}\log\frac{\sigma^2}{D}+\xi_n+o(\xi_n)\bigg)\label{mddconverse00},
\end{align}
we have
\begin{align}
\rmP_{\rme,n}(M)
&\geq \left(1-\frac{1}{\sqrt{n}}\right)\Pr\{Z\in\calP'\cap\calQ\}\label{mddconverse01}.
\end{align}
Using the moderate deviations theorem in \cite[Theorem 3.7.1]{dembo2009large} and the definition of $\calP'$ in \eqref{def:calp'n}, we obtain that
\begin{align}
\Pr\{Z\in\calP'\}
&=\Pr\bigg\{\frac{1}{n}\sum_{i=1}^nX_i^2>\sigma^2+\xi_n\bigg\}-\Pr\bigg\{\frac{1}{n}\sum_{i=1}^nX_i^2>\sigma^2+2\xi_n\bigg\}\\
&=\exp\bigg\{-\frac{n\xi_n^2}{2\rmV(\sigma^2,\zeta)}+o(n\xi_n^2)\bigg\}-\exp\bigg\{-\frac{4n\xi_n^2}{2\rmV(\sigma^2,\zeta)}+o(n\xi_n^2)\bigg\}\label{mddconverse02}.
\end{align} 
Invoking Lemma \ref{largediid}, using the definition of $\calQ$ in \eqref{def:calq} and the definition of $P_Y=\sigma^2-D$, we have
\begin{align}
\Pr\{Z\notin\calQ\}
&=\Pr\bigg\{\frac{1}{n}\sum_{i=1}^nX_i^2<D-P_Y\bigg\}\\
&=\Pr\bigg\{\frac{1}{n}\sum_{i=1}^nX_i^2<\sigma^2-2P_Y\bigg\}\\
&\leq \exp(-nt_2)\label{mddconverse03}
\end{align}
for some $t_2>0$.

Combining the results in \eqref{mddconverse01} to \eqref{mddconverse03} and using the inequality that $\Pr\{Z\in\calP'\cap\calQ\}\geq \Pr\{Z\in\calP'\}-\Pr\{Z\in\calQ\}$, for any $(n,M)$-code satisfying \eqref{mddconverse00}, we have
\begin{align}
\frac{\rmP_{\rme,n}(M)}{1-\frac{1}{\sqrt{n}}}
&\geq \exp\bigg\{-\frac{n\xi_n^2}{2\rmV(\sigma^2,\zeta)}+o(n\xi_n^2)\bigg\}-\exp\bigg\{-\frac{4n\xi_n^2}{2\rmV(\sigma^2,\zeta)}+o(n\xi_n^2)\bigg\}-\exp\{-nt_2\}\label{mddconverse0},
\end{align}
for some $t_2>0$. Note that the first term on the right hand side of~\eqref{mddconverse0} dominates as $n\to\infty$. From the results in \eqref{mddconverse00} and~\eqref{mddconverse0}, we conclude that 
\begin{align}
\liminf_{n\to\infty}-\frac{1}{n\xi_n^2}\log\rmP_{\rme,n}(\exp(n (R(\sigma^2,D) +\xi_n) )
&\leq \frac{1}{2\rmV(\sigma^2,\zeta)}
\end{align}

\subsection{Proof for  the I.I.D. Gaussian Codebook}
The proof for i.i.d.\ Gaussian codebooks is similar to the proof in Section \ref{sec:iidach} and \ref{sec:iidconverse} with the use of Lemma \ref{largediid} and \cite[Theorem 3.7.1]{dembo2009large} as in Sections \ref{sec:proofmddsphere} and \ref{sec:proofmddsphere_conv} and is thus omitted.

\section{Proof of Large Deviation Asymptotics (Theorem \ref{ldresult})}
\label{sec:proofld}

\subsection{Preliminaries}

The following properties of the quantities $R_{\rm{sp}}(z)$ in \eqref{def:rspa2}, $R_{\rm{iid}}(s^*(z),z)$ (cf. \eqref{def:riidsapha} and \eqref{def:sstar}) and $\Lambda_{X^2}^*(t)$ are useful in the proof of Theorem \ref{ldresult}.
\begin{lemma}
\label{property}
The following  claims hold.
\begin{enumerate}
\item Concerning $R_{\rm{sp}}(z)$,

\begin{itemize}
\item[(a)] $R_{\rm{sp}}(z)$ is increasing in $z$ if $z\geq |\sigma^2-2D|$;
\item[(b)]  $R_{\rm{sp}}(\sigma^2)=\frac{1}{2}\log\frac{\sigma^2}{D}$, $\lim_{z\to r_2^2} R_{\rm{sp}}(z)=\infty$.
\end{itemize}

\item Concerning $R_{\rm{iid}}(s^*(z),z)$,
\begin{itemize}
\item[(a)] $s^*(z)>0$ if and only if $z>\max(0,2D-\sigma^2)$;
\item[(b)] $R_{\rm{iid}}(s^*(z),z)=0$ if $z=\max\{0,2D-\sigma^2\}$ and $R_{\rm{iid}}(s^*(\sigma^2),\sigma^2)=\frac{1}{2}\log\frac{\sigma^2}{D}$;
\item[(c)] $R_{\rm{iid}}(s^*(z),z)=\sup_{s\geq 0} R_{\rm{iid}}(s,z)$ and thus 
$R_{\rm{iid}}(s^*(z),z)$ is increasing in $z$ for $z\geq \max\{0,2D-\sigma^2\}$.
\end{itemize}

\item Concerning $\Lambda_{X^2}^*(t)$ (cf. \cite[Chapter 3]{dembo2009large}),
\begin{itemize}
\item[(a)] $\Lambda_{X^2}^*(t)$ is convex and non-decreasing in $t$ for $t\geq 0$;
\item[(b)] $\Lambda_{X^2}^*(t)=0$ if $t\leq \sigma^2$;
\item[(c)] $\Lambda_{X^2}^*(t)$ is increasing in $t$ for $t\geq \sigma^2$.
\end{itemize}
\end{enumerate}
\end{lemma}
The proof of Lemma \ref{property} is omitted since it follows either from simple algebra or from \cite[Chapter 3]{dembo2009large}.

Furthermore, we have the following lemma concerning an important property of the function $\underline{g}(n,z)$ (cf. \eqref{def:undergnz}), which plays an important role in proving the ensemble tight excess-distortion exponent for spherical codebooks.
\begin{lemma}
\label{propugnz}
For any $\alpha\in[\sigma^2,r_2^2)$, there exists a unique $\beta\in(r_1^2,|\sigma^2-2D|)$ such that
\begin{align}
\underline{g}(n,\beta)&=\underline{g}(n,\alpha)\label{gp1},\\
\alpha+\beta&\leq 2\sigma^2\label{gp2}.
\end{align}
\end{lemma}
The proof of Lemma \ref{propugnz} is given in Appendix \ref{proofpropugnz}.

\subsection{Achievability Proof for  the spherical Codebook}
Recall that $P_Y=\sigma^2-D$. Invoking the definitions of $r_1$ in \eqref{def:r1} and $\underline{g}(\cdot)$ in \eqref{def:undergnz}, we conclude that $r_1^2\leq |\sigma^2-2D|<\sigma^2$, $\underline{g}(n,z)$ is decreasing in $z$ if $z\in(|\sigma^2-2D|,r_2^2)$ and $\underline{g}(n,z)$ is increasing in $z$ if $z\in(r_1^2,|\sigma^2-2D|)$

Using the expression for the excess-distortion probability in \eqref{pe4sphere}, given any $\alpha$ such that $\alpha\in[\sigma^2,r_2^2)$, we can upper bound $\rmP_{\rme,n}(M)$ as follows
\begin{align}
\rmP_{\rme,n}(M)
&\leq \Pr\bigg\{\frac{1}{n}\sum_{i=1}^n X_i^2<r_1^2\bigg\}+\Pr\bigg\{\frac{1}{n}\sum_{i=1}^n X_i^2>r_2^2\bigg\}+\int_{r_1^2}^{r_2^2} (1-\Psi(n,z))^M f_Z(z)\, \rmd z\\
\nn&\leq \Pr\bigg\{\frac{1}{n}\sum_{i=1}^n X_i^2<\beta\bigg\}+\Pr\bigg\{\frac{1}{n}\sum_{i=1}^n X_i^2>\alpha\bigg\}+\int_{\beta}^{|\sigma^2-2D|}(1-\underline{g}(n,z))^M f_Z(z)\, \rmd z\\
&\qquad+\int_{|\sigma^2-2D|}^{\alpha}(1-\underline{g}(n,z))^M f_Z(z)\, \rmd z\label{verify-1}\\
&\leq \Pr\bigg\{\frac{1}{n}\sum_{i=1}^n (X_i^2-\sigma^2)<\beta-\sigma^2\bigg\}+\Pr\bigg\{\frac{1}{n}\sum_{i=1}^n X_i^2>\alpha\bigg\}+(1-\underline{g}(n,\alpha))^M\label{verify0}\\
&\leq \Pr\bigg\{\frac{1}{n}\sum_{i=1}^n (X_i^2-\sigma^2)>\sigma^2-\beta\}\bigg\}+\Pr\bigg\{\frac{1}{n}\sum_{i=1}^n (X_i^2-\sigma^2)>\alpha-\sigma^2\bigg\}+\exp\{-M\underline{g}(n,\alpha)\}\label{verify1}\\
&\leq 2\Pr\bigg\{\frac{1}{n}\sum_{i=1}^n X_i^2>\alpha\bigg\}+\exp\{-M\underline{g}(n,\alpha)\}\label{verify2},
\end{align} 
where \eqref{verify-1} follows from the result in Lemma \ref{propugnz} which states that there exists a unique $\beta\in(r_1^2,|\sigma^2-2D|)$ such that $\underline{g}(n,\beta)=\underline{g}(n,\alpha)$ for any $\alpha\in[\sigma^2,r_2^2)$; \eqref{verify0} follows since i) $\alpha\geq \sigma^2>|\sigma^2-2D|$ and $\underline{g}(n,z)$ is decreasing in $z$ for $z\geq |\sigma^2-2D|$ and , ii) $\underline{g}(n,z)$ is increasing in $z$ for $z\in(r_1^2,|\sigma^2-2D|)$; and iii) the result in Lemma \ref{propugnz} which states that $\underline{g}(n,\beta)=\underline{g}(n,\alpha)$ (cf. \eqref{gp1}) and $\beta\in(r_1^2,|\sigma^2-2D|)$; \eqref{verify1} follows since $(1-a)^M\leq \exp\{-Ma\}$ for any $a\in[0,1)$; and \eqref{verify2} follows since $\sigma^2-\beta\geq \alpha-\sigma^2$, which is implied by \eqref{gp2} in Lemma \ref{propugnz}.

Now, given any positive $\delta\in(0,1)$, recalling the definition of $R_{\rm{sp}}(\cdot)$ in \eqref{def:rspa2}, we choose $M$ such that
\begin{align}
\log M=(1+\delta)(n-1)R_{\rm{sp}}(\alpha)+\log\frac{\sqrt{\pi}n\Gamma(\frac{n+1}{2})}{\Gamma(\frac{n+2}{2})}\label{eechoosemach}.
\end{align}
Using the definitions of $R_{\rm{sp}}(\cdot)$ in \eqref{def:rspa2} and $\underline{g}(n,z)$ in \eqref{def:undergnz}, we obtain that 
\begin{align}
\exp\{-M\underline{g}(n,\alpha)\}
&= \exp\big\{-\exp\{(n-1)\delta R_{\rm{sp}}(\alpha)\}\big\}\label{eesphereach2},
\end{align}

which vanishes doubly exponentially fast for $\alpha\geq \sigma^2$. Invoking Cram\'er's Theorem~\cite[Theorem 2.2.3]{dembo2009large} and the definition of $\Lambda_{X^2}^*(\cdot)$, we obtain that
\begin{align}
\Pr\bigg\{\frac{1}{n}\sum_{i=1}^nX_i^2>\alpha\bigg\}
&\leq \exp\big\{-n\Lambda_{X^2}^*(\alpha)\big\}\label{eesphereach3}.
\end{align}
Therefore, using \eqref{verify2} to \eqref{eesphereach3}, noting that $ {\sqrt{\pi}n\Gamma(\frac{n+1}{2})}/{\Gamma(\frac{n+2}{2})}=\Theta(1/\sqrt{n})$, recalling that $R_{\rm{sp}}(z)$ is increasing in $z$ for $z\geq \sigma^2$ (cf.~Claim~(i)(a) in Lemma \ref{property}), using the result that $R_{\rm{sp}}(\sigma^2)=\frac{1}{2}\log\frac{\sigma^2}{D}$ (cf.~Claim~(i)(b) in Lemma \ref{property}) and letting $\delta\downarrow 0$, we conclude that for all $\alpha\in[\sigma^2,r_2^2)$, 
\begin{align}
\liminf_{n\to\infty}-\frac{1}{n}\log \rmP_{\rme,n}(\lceil \exp(nR)\rceil)&\geq \Lambda_{X^2}^*(\alpha),
\end{align}
where where $\alpha$ is determined from $R=R_{\rm{sp}}(\alpha)$ (cf. \eqref{def:rspa2}).

The proof for $R\in[0,\frac{1}{2}\log\frac{\sigma^2}{D})$ follows trivially by noting that any $(n,M)$-code satisfies that $\rmP_{\rme,n}(M)\leq 1$.

\subsection{Ensemble Converse Proof for the  spherical Codebook}

Fix any $\alpha$ such that $\alpha\in[\sigma^2,r_2^2)$ (cf. \eqref{def:r2} for the definition of $r_2$). Let
\begin{align}
\tilde{\calP}
&:=\{r\in\bbR:\alpha\leq r<r_2^2\},\\
\tilde{\calQ}
&:=\{r\in\bbR:r-|\sigma^2-2D|\geq 0\}.
\end{align}
Note that $r\in\tilde{\calQ}$ implies that $r+(\sigma^2-2D)\geq 0$. 

Using the result in \eqref{peforboth} and the definition of $\overline{g}(\cdot)$ in \eqref{def:overlineg}, we conclude that for sufficiently large $n$,
\begin{align}
\rmP_{\rme,n}(M)
&\geq \int_{\alpha}^{r_2^2}(1-\overline{g}(n,z))^M 1\{z\in\tilde{\calP}\cap\tilde{\calQ}\}f_Z(z)\, \rmd z\\
&\geq (1-\overline{g}(n,\alpha))^M\Pr\bigg\{\frac{1}{n}\sum_{i=1}^nX_i^2\in\tilde{\calP}\cap\tilde{\calQ}\bigg\}\label{useognonincrease}\\
&\geq \exp\bigg\{-M\frac{\overline{g}(n,\alpha)}{1-\overline{g}(n,\alpha)}\bigg\}\Pr\bigg\{\frac{1}{n}\sum_{i=1}^nX_i^2\in\tilde{\calP}\cap\tilde{\calQ}\bigg\}\label{uselblog2}\\
&\geq \exp\big\{-2M\overline{g}(n,\alpha)\big\}\Pr\bigg\{\frac{1}{n}\sum_{i=1}^nX_i^2\in\tilde{\calP}\cap\tilde{\calQ}\bigg\}\label{useogsmallnlarge},
\end{align}
where \eqref{useognonincrease} follows since $\overline{g}(n,z)$ is decreasing in $z$ for $z\geq |\sigma^2-2D|$ and $\alpha\geq \sigma^2>|\sigma^2-2D|$; \eqref{uselblog2} follows since $(1-a)^M\geq \exp\{-M\frac{a}{1-a}\}$ for any $a\in[0,1)$; and \eqref{useogsmallnlarge} follows since $\overline{g}(n,\alpha)\leq \frac{1}{2}$ for $n$ sufficiently large.

For any $M$ such that
\begin{align}
\log M&\leq -\log \overline{g}(n,\alpha)-\log 2-\frac{1}{2}\log n\label{sphereeecons1},
\end{align}
using \eqref{useogsmallnlarge} and the inequality that $\exp\{-a\}\geq 1-a$, we have that for sufficiently large $n$,
\begin{align}
\rmP_{\rme,n}(M)
&\geq \bigg( 1-\frac{1}{\sqrt{n}}\bigg)\Pr\bigg\{\frac{1}{n}\sum_{i=1}^nX_i^2\in\tilde{\calP}\cap\tilde{\calQ}\bigg\}\label{sphereeecons2}.
\end{align}

Note that
\begin{align}
& \Pr\bigg\{\frac{1}{n}\sum_{i=1}^nX_i^2\in\tilde{\calP}\cap\tilde{\calQ}\bigg\} \nn\\*
 &=\Pr\bigg\{\max\{\alpha,|\sigma^2-2D|\}\leq \frac{1}{n}\sum_{i=1}^nX_i^2<r_2^2\bigg\}\\*
&=\Pr\bigg\{\frac{1}{n}\sum_{i=1}^nX_i^2\geq \alpha  \bigg\}-\Pr\bigg\{\frac{1}{n}\sum_{i=1}^nX_i^2\geq r_2^2\bigg\},
\end{align}
where  the final equality holds because  $\alpha^2$ is chosen to be in $ [\sigma^2,r_2^2)$  so $\alpha^2\geq \sigma^2>|\sigma^2-2D|$.  
Invoking Cram\'er's theorem~\cite[Theorem 2.2.3]{dembo2009large}, we obtain that for sufficiently large $n$ and any positive number $\delta\in(0,1)$, 
\begin{align}
\nn&\Pr\bigg\{\frac{1}{n}\sum_{i=1}^nX_i^2\in\tilde{\calP}\cap\tilde{\calQ}\bigg\}\\
&\geq \exp\bigg\{-n(1+\delta)\Lambda_{X^2}^*(\alpha)\bigg\}-\exp\bigg\{-n(1+\delta)\Lambda_{X^2}^*(r_2^2)\bigg\}\\
&\geq \frac{1}{2}\exp\bigg\{-n(1+\delta)\Lambda_{X^2}^*(\alpha)\bigg\}\label{sphereeecons3},
\end{align}
where \eqref{sphereeecons3} holds since $\sigma^2\leq \alpha<r_2^2$ and $\Lambda_{X^2}^*(t)$ is increasing in $t$ for all $t\geq \sigma^2$ (cf. Claim (iii)(c) in Lemma \ref{property}).

Using the definitions of $\overline{g}(\cdot)$ in \eqref{def:overlineg}, $R_{\rm{sp}}(\cdot)$ in \eqref{def:rspa2}, invoking the bounds in \eqref{sphereeecons1}, \eqref{sphereeecons2}, \eqref{sphereeecons3}, recall thating $R_{\rm{sp}}(z)$ is increasing in $z$ for $z\geq \sigma^2$ (cf.~Claim (i)(a) in Lemma \ref{property}) and $R_{\rm{sp}}(\sigma^2)=\frac{1}{2}\log\frac{\sigma^2}{D}$ (cf.~Claim (i)(b) in Lemma \ref{property}) and letting $\delta\to 0$, we conclude that for any $\alpha\in[\sigma^2,r_2^2)$
\begin{align}
\liminf_{n\to\infty}-\frac{1}{n}\log\rmP_{\rme,n}(\lceil \exp(nR)\rceil)
&\leq \Lambda_{X^2}^*(\alpha).
\end{align}
where $\alpha$ is determined from $R=R_{\rm{sp}}(\alpha)$.

\subsection{Achievability Proof for the  I.I.D.\ Gaussian  Codebook}
Fix $\alpha$ such that $\alpha>\max\{0,2D-\sigma^2\}$. Invoking the conclusion in \eqref{iidach1}, for any $x^n$ such that $\frac{1}{n } {\|x^n\|^2} \leq \alpha$, we have that for sufficiently large $n$ and any positive $\delta$,
\begin{align}
\Upsilon\bigg( n, \frac{1}{n}\|x^n\|^2\bigg)
&\geq \Upsilon(n,\alpha)\\
&\geq \exp\{-n(1+\delta)R_{\rm{iid}}(s^*(\alpha),\alpha)\}\label{ldiidnonincreasing}.
\end{align}
Invoking \eqref{peforboth}, the excess-distortion probability can be upper bounded as follows
\begin{align}
\rmP_{\rme,n}(M)
&=\bbE_{X^n}\Big[(1-\Pr\{d(X^n,Y^n)\leq D|X^n\})^M\Big]\label{ce2}\\
&=\int_0^\infty (1-\Upsilon(n,z))^Mf_Z(z)\, \rmd z\label{usedefUp}\\
&\leq\int_0^{\alpha} (1-\Upsilon(n,z))^Mf_Z(z)\, \rmd z+\Pr\bigg\{\frac{1}{n}\sum_{i=1}^n X_i^2>\alpha\bigg\}\label{useUple1}\\
&\leq \int_0^{\alpha} \big(1-\exp\{-n(1+\delta)R_{\rm{iid}}(s^*(\alpha),\alpha)\}\big)^M f_Z(z)\, \rmd z+\Pr\bigg\{\frac{1}{n}\sum_{i=1}^n X_i^2>\alpha\bigg\}\label{useldiid}\\
&\leq \exp\big\{-M\exp\{-n(1+\delta)R_{\rm{iid}}(s^*(\alpha),\alpha)\}\big\}+\Pr\bigg\{\frac{1}{n}\sum_{i=1}^n X_i^2>\alpha\bigg\}\label{eeiidach1},
\end{align}
where \eqref{usedefUp} follows from the definition of $\Upsilon(\cdot)$ in \eqref{def:upsilonnz}; \eqref{useUple1} follows since $\Upsilon(n,z)\geq 0$; \eqref{useldiid} follows from~\eqref{ldiidnonincreasing};  and~\eqref{eeiidach1} follows since $(1-a)^M\leq \exp\{-Ma\}$ for any $a\in[0,1)$.

Recall the definitions of $s^*(\cdot)$ in \eqref{def:sstar} and $R_{\rm{iid}}(\cdot)$ in \eqref{def:riidsapha}. Choose $M$ such that
\begin{align}
\log M=n(1+2\delta)R_{\rm{iid}}(s^*(\alpha),\alpha)\label{iideeachrate}.
\end{align}
Invoking the definition of $\Lambda_{X^2}^*(\cdot)$ (cf. Section \ref{sec:notation}), the conclusion in \eqref{eeiidach1} and Cram\'er's Theorem~\cite[Theorem 2.2.3]{dembo2009large}, we conclude that for sufficiently large $n$ and arbitrary positive $\delta$,
\begin{align}
\rmP_{\rme,n}(M)
&\leq \exp\big\{-\exp\big\{n\delta R_{\rm{iid}}(s^*(\alpha),\alpha)\big\}\big\}+\exp\big\{-n\Lambda_{X^2}^*(\alpha)\big\}\label{iideeachexp}.
\end{align}
Recall that $R_{\rm{iid}}(s^*(\sigma^2),\sigma^2)=\frac{1}{2}\log\frac{\sigma^2}{D}$ and $R_{\rm{iid}}(s^*(z),z)$ is positive and increasing for $z>\max(0,2D-\sigma^2)$ (cf.~Claim (ii) in Lemma \ref{property}.  Thus, the first term in \eqref{iideeachexp} vanishes doubly exponentially fast since $\alpha>\max(0,2D-\sigma^2)$. Using the results in \eqref{iideeachrate} and~\eqref{iideeachexp}, noting the fact that $\Lambda_{X^2}^*(t)=0$ if $t\leq \sigma^2$ (cf. Claim (iii)(b) in Lemma \ref{property}) and letting $\delta\downarrow 0$, we conclude that for any $\alpha>\max\{0,2D-\sigma^2\}$,
\begin{align}
\liminf_{n\to\infty}-\frac{1}{n}\log \rmP_{\rme,n}(\lceil \exp(nR)\rceil)&\geq \Lambda_{X^2}^*(\alpha),
\end{align}
where $R$ is determined from $R=R_{\rm{iid}}(s^*(\alpha),\alpha)$ (cf.~\eqref{def:riidsapha}).

\subsection{Ensemble Converse Proof for the  I.I.D.\ Gaussian   Codebook}
Fix any $\alpha$ such that $\alpha>\max\{0,2D-\sigma^2\}$. Using the  strong large deviations result in \eqref{iidach1}, we conclude that for $n$ large enough and any positive number $\delta\in(0,1)$, given any $x^n$ such that $\frac{1}{n} {\|x^n\|^2}\geq \alpha$,
\begin{align}
\Upsilon\bigg( n,\frac{1}{n}\|x^n\|^2\bigg)
&\leq \Upsilon(n,\alpha)\\
&\leq \exp\{-n(1-\delta)R_{\rm{iid}}(s^*(\alpha),\alpha)\}\label{iidldconverse}.
\end{align}
From \eqref{peforboth} and \eqref{iidldconverse}, we conclude that for sufficiently large $n$,
\begin{align}
\rmP_{\rme,n}(M)
&=\int_0^\infty (1-\Upsilon(n,z))^Mf_Z(z)\, \rmd z\\
&\geq \int_{\alpha}^\infty (1-\Upsilon(n,z))^Mf_Z(z)\, \rmd z\\
&\geq \int_{\alpha}^\infty\big(1-\exp\big\{-n(1-\delta)R_{\rm{iid}}(s^*(\alpha),\alpha)\big\}\big)^M f_Z(z)\, \rmd z\\
&\geq \exp\big\{-2M\exp\big\{-n(1-\delta)R_{\rm{iid}}(s^*(\alpha),\alpha)\big\}\big\}\Pr\bigg\{\frac{1}{n}\sum_{i=1}^n X_i^2>\alpha\bigg\}\label{eeiidac1},
\end{align}
where \eqref{eeiidac1} follows because i) $(1-a)^M\geq \exp\{-M\frac{a}{1-a}\}$ and ii) for $n$ sufficiently large, $\exp\big\{-n(1-\delta)R_{\rm{iid}}(s^*(\alpha),\alpha)\big\}\leq \frac{1}{2}$. Using the bound in \eqref{eeiidac1} and Cram\'er's theorem~\cite[Theorem 2.2.3]{dembo2009large}, we conclude that for any $\alpha>\max(0,2D-\sigma^2)$, if $M$ is chosen such that
\begin{align}
\log M\leq n(1-\delta)R_{\rm{iid}}(s^*(\alpha),\alpha)-\log n-\log 2,
\end{align}
then for sufficiently large $n$, using the inequality $\exp(-a)\geq 1-a$ for $a\in[0,1)$, we obtain
\begin{align}
\rmP_{\rme,n}(M)
&\geq \bigg( 1-\frac{1}{n}\bigg)\exp\big\{-n(1+\delta)\Lambda_{X^2}^*(\alpha)\big\}.
\end{align}
Hence, given any $\alpha>\max(0,2D-\sigma^2)$, recalling that $R_{\rm{iid}}(s^*(\sigma^2),\sigma^2)=\frac{1}{2}\log\frac{\sigma^2}{D}$ and $R_{\rm{iid}}(s^*(z),z)$ is positive and increasing for $z>\max(0,2D-\sigma^2)$ (cf.~Claim~(ii) in Lemma \ref{property}), using the fact that $\Lambda_{X^2}^*(t)=0$ if $t\leq \sigma^2$ (cf.~Claim~(iii)(b) in Lemma \ref{property}) and letting $\delta\downarrow 0$, we have shown that for any seqeunce of $(n,M)$-codes such that
\begin{align}
\liminf_{n\to\infty}-\frac{1}{n}\log \rmP_{\rme,n}(\lceil \exp(nR)\rceil)&\leq \Lambda_{X^2}^*(\alpha),
\end{align}
where $\alpha$ is determined from $R=R_{\rm{iid}}(s^*(\alpha),\alpha)$.

\appendix
\subsection{Proof of Lemma \ref{concentrate}}
\label{proof:concentrate}
Note that $T$ (cf. \eqref{def:T}) is finite since $\bbE[X^6]$ is finite~(cf. \eqref{sourceconstraint}). Using the definitions of $a_n$ in \eqref{def:an}, $b_n$ in \eqref{def:bn}, $\calP$ in \eqref{def:calp}, and the Berry-Esseen theorem, we obtain  
\begin{align}
\Pr\{Z\in\calP\}
&=\Pr\bigg\{\frac{1}{n}\sum_{i=1}^n X_i^2>\sigma^2+b_n\bigg\}-\Pr\bigg\{\frac{1}{n}\sum_{i=1}^n X_i^2>\sigma^2+a_n\bigg\}\\
&=\Pr\bigg\{\frac{1}{n}\sum_{i=1}^n (X_i^2-\sigma^2)>\sqrt{\frac{\rmV}{n}}\rmQ^{-1}(\varepsilon)\bigg\}-\Pr\bigg\{\frac{1}{n}\sum_{i=1}^n (X_i^2-\sigma^2)>\sqrt{\rmV\frac{\log n}{n}}\bigg\}\\
&\geq \varepsilon-\frac{6T}{\sqrt{n}\, \rmV^{3/2}}-\bigg( \rmQ(\sqrt{\log n})+\frac{6T}{\sqrt{n}\, \rmV^{3/2}}\bigg)\\
&\geq \varepsilon+O\bigg( \frac{1}{\sqrt{n}}\bigg)\label{upprmq},
\end{align}
where \eqref{upprmq} follows since $\rmQ(x)\leq \exp\{-\frac{x^2}{2}\}$ and $T$ is finite.
Similarly, using the definition of $\calQ$ in \eqref{def:calq} and the Berry-Esseen theorem, we obtain  
\begin{align}
\Pr\{Z\notin\calQ\}
&=\Pr\bigg\{\frac{1}{n}\sum_{i=1}^n X_i^2+P_Y-D\leq 0\bigg\}\\
&=\Pr\bigg\{\frac{1}{n}\sum_{i=1}^n X_i^2\leq \sigma^2+(D-P_Y-\sigma^2)\bigg\}\\
&=\Pr\bigg\{\frac{1}{n}\sum_{i=1}^n X_i^2\leq \sigma^2-2P_Y\bigg\}\\
&\leq \rmQ\bigg( 2\sqrt{\frac{n}{\rmV}}P_Y\bigg)+\frac{6T}{\sqrt{n}\, \rmV^{3/2}}\\
&\leq \exp\bigg\{-\frac{2nP_Y}{\rmV}\bigg\}+\frac{6T}{\sqrt{n}\, \rmV^{3/2}}\label{chooseeta0}\\
&=O\bigg( \frac{1}{\sqrt{n}}\bigg)\label{chooseeta},
\end{align}
where \eqref{chooseeta} follows since the first term in \eqref{chooseeta0} vanishes exponentially fast and is thus dominated by the second term. Combining \eqref{upprmq} and \eqref{chooseeta}, we have
\begin{align}
\Pr\{Z\in\calP\cap\calQ\}
&\geq \Pr\{Z\in\calP\}-\Pr\{Z\notin\calQ\}\\*
&\geq \varepsilon+O\bigg( \frac{1}{\sqrt{n}}\bigg).
\end{align}

\subsection{Proof of Lemma \ref{comld}}
\label{proofcomld}
For simplicity, in the proof of Lemma \ref{comld}, for any $z\in\bbR_+$, we let
\begin{align}
R_{\rm{iid}}^*(z)&:=R_{\rm{iid}}(s^*(z),z).
\end{align}
Recalling the results in \eqref{boundsldsphere} and \eqref{ree:iid}, we  find that the ensemble excess-distortion exponents of the spherical and i.i.d Gaussian codebooks for a given rate $R$ are determined by the corresponding parameters $\alpha$ in $\Lambda_{X^2}^*(\alpha)$. Using conclusions (i) and (ii) regarding the increasing properties of $R_{\rm{sp}}(z)$ and $R_{\rm{iid}}^*(z)$ in Lemma \ref{property}, we obtain that for any given $R>\frac{1}{2}\log\frac{\sigma^2}{D}$, there exists unique $\alpha_{\rm{sp}}(R)$ and $\alpha_{\rm{iid}}(R)$ such that,
\begin{align}
R_{\rm{sp}}(\alpha_{\rm{sp}}(R))&=R,\\
R_{\rm{iid}}^*(\alpha_{\rm{iid}}(R))&=R\label{def:alphaiidR}.
\end{align}
From conclusion (iii) in Lemma \ref{property}, we know that $\Lambda_{X^2}^*(t)$ is an increasing function of $t$ for $t\geq \sigma^2$. Thus, to prove Lemma \ref{comld}, it suffices to show that given any rate $R>\frac{1}{2}\log\frac{\sigma^2}{D}$, we have
\begin{align}
\alpha_{\rm{sp}}(R)<\alpha_{\rm{iid}}(R)\label{eqn:alt}.
\end{align}
For any $R\geq R_{\rm{iid}}^*(r_2^2)$, from \eqref{def:alphaiidR} and the fact that $R_{\rm{iid}}^*(z)$ is increasing in $z$ for $z\geq \sigma^2$ (cf. Claim (ii) in Lemma \ref{property}), we have
\begin{align}
\alpha_{\rm{iid}}(R)&\geq r_2^2\geq \alpha_{\rm{sp}}(R),
\end{align}
since $\alpha_{\rm{sp}}(R)<r_2^2$ for any $R\in[\sigma^2,\infty)$ (cf. \eqref{def:rspa2}).

\begin{figure}[t]
\centering
\includegraphics[width=9cm]{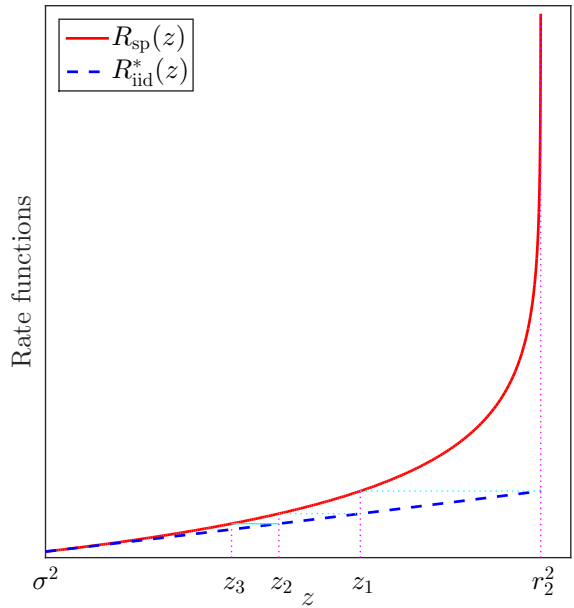}
\caption{Illustration of the proof idea for \eqref{eqn:alt2}.}
\label{comparers}
\end{figure}

In the following, we will show that \eqref{eqn:alt} also holds for any $R\in(\frac{1}{2}\log\frac{\sigma^2}{D},R_{\rm{iid}}^*(r_2^2))$. Recall that both $R_{\rm{sp}}(z)$ and $R_{\rm{iid}}(z)$ are increasing in $z$ for $z\in(\sigma^2,r_2^2)$ (cf. Lemma \ref{property}). Thus, to prove \eqref{eqn:alt} for $R\in(\frac{1}{2}\log\frac{\sigma^2}{D},R_{\rm{iid}}^*(r_2^2))$ is equivalent to show that for any $z\in(\sigma^2,r_2^2)$,
\begin{align}
R_{\rm{sp}}(z)>R_{\rm{iid}}^*(z)\label{eqn:alt2}.
\end{align}
From Lemma \ref{property}, we have that 
\begin{align}
\lim_{z\to r_2^2}R_{\rm{sp}}(z)&=\infty>R_{\rm{iid}}^*(r_2^2).
\end{align}
Therefore, there exists $z_1\in(\sigma^2,r_2^2)$ such that
\begin{align}
R_{\rm{sp}}(z_1)=R_{\rm{iid}}^*(r_2^2).
\end{align}
Using the increasing nature of $R_{\rm{sp}}(z)$ and $R_{\rm{iid}}^*(z)$, we conclude that for any $z\in[z_1,r_2^2)$, we have
\begin{align}
R_{\rm{sp}}(z)>R_{\rm{sp}}(z_1)=R_{\rm{iid}}^*(r_2^2)>R_{\rm{iid}}^*(z).
\end{align}
Note that $R_{\rm{sp}}(z_1)>R_{\rm{iid}}^*(z_1)$. Thus, there exists $z_2\in(\sigma^2,z_1)$ such that for any $z\in[z_2,z_1)$, we have
\begin{align}
R_{\rm{sp}}(z)>R_{\rm{sp}}(z_2)=R_{\rm{iid}}^*(z_1)>R_{\rm{iid}}^*(z).
\end{align}
Similarly, we can show that there exists a sequence $\{z_i\}_{i\geq 3}$ such that $z_i\in(\sigma^2,z_{i-1})$, $\lim_{i\to\infty}z_i=\sigma^2$, and for each $z\in[z_i,z_{i-1})$, we have
\begin{align}
R_{\rm{sp}}(z)>R_{\rm{sp}}(z_i)=R_{\rm{iid}}^*(z_{i-1})>R_{\rm{iid}}^*(z).
\end{align}
We illustrate the proof in Figure \ref{comparers}.

\subsection{Proof of Lemma \ref{propugnz}}
\label{proofpropugnz}
\begin{figure}[t]
\centering
\includegraphics[width=9cm]{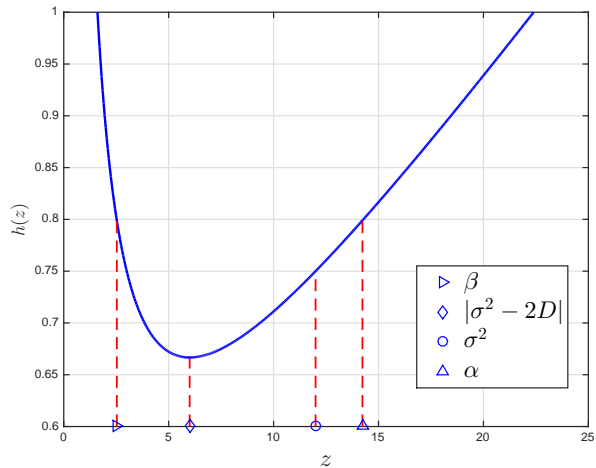}
\caption{Plot of $h(z)$.}
\label{plothz}
\end{figure}

Recall that $P_Y=\sigma^2-D$. Given any $z\in\bbR_+$, let
\begin{align}
h(z):=\frac{(z+P_Y-D)^2}{4zP_Y}.
\end{align}
Recall the definition of $\underline{g}(n,z)$ in \eqref{def:undergnz}, we conclude that it suffices to show that for any $\alpha\in[\sigma^2,r_2^2)$, there exists a unique $\beta\in(r_1^2,|\sigma^2-2D|)$ such that
\begin{align}
h(\alpha)=h(\beta),\quad\mbox{and}\quad \alpha+\beta\leq 2\sigma^2.
\end{align}
We obtain that
\begin{align}
\frac{\partial h(z)}{\partial z}&=\frac{z^2-(\sigma^2-2D)^2}{4z^2(\sigma^2-D)},\label{ch1}\\
h(r_1^2)&=h(r_2^2)=1\label{ch2}.
\end{align}
Note that \eqref{ch1} implies that the function $h(z)$ is increasing in $z\in[|\sigma^2-2D|,r_2^2)$ and decreasing  in $z\in(r_1^2,|\sigma^2-2D|)$. Hence, using \eqref{ch2}, we can conclude that, given any $\alpha\in[\sigma^2,r_2^2]\subseteq[|\sigma^2-2D|,r_2^2)$, there exists a unique $\beta\in(r_1^2,|\sigma^2-2D|)$ such that $h(\alpha)=h(\beta)$. Thus, in the following, we only need to show that $\beta\leq 2\sigma^2-\alpha$. By noticing that $h(z)\leq h(\alpha)=h(\beta)$ only for all $z\in[\alpha,\beta]$, we conclude that it suffices to show that $h(2\sigma^2-\alpha)\leq h(\alpha)$. To illustrate our arguments here, we plot $h(z)$ in Figure \ref{plothz} for the case where $\sigma^2=12$ and $D=3$.

For any $\alpha\in[\sigma^2,r_2^2)$, we find that
\begin{align}
h(\alpha)-h(2\sigma^2-\alpha)
&=\frac{(\alpha-\sigma^2)((\alpha-\sigma^2)^2+4D(D-\sigma^2))}{2\alpha(\sigma^2-D)(\alpha-2\sigma^2)}\label{h1p}.
\end{align}
Since $\alpha\in[\sigma^2,r_2^2)$, using the definitions of $r_1$ in \eqref{def:r1} and $r_2$ in \eqref{def:r2}, we have $r_2^2<r_1^2+r_2^2=2\sigma^2$ and thus $\alpha<2\sigma^2$. Recalling the fact that $\sigma^2>D$, we conclude that
\begin{align}
2\alpha(\sigma^2-D)(\alpha-2\sigma^2)<0\label{h2p}.
\end{align}
Furthermore, for any $\alpha\in[\sigma^2,r_2^2)$, we have $\alpha-\sigma^2\geq 0$.
\begin{align}
(\alpha-\sigma^2)^2+4D(D-\sigma^2)
&<(r_2^2-\sigma^2)^2+4D(D-\sigma^2)=0\label{h3p}.
\end{align}
Therefore, combining \eqref{h1p}, \eqref{h2p} and \eqref{h3p}, we conclude that $h(2\sigma^2-\alpha)\leq h(\alpha)$. This completes the proof of Lemma~\ref{propugnz}.

\bibliographystyle{IEEEtran}
\bibliography{../IEEEfull_lin}

\end{document}